\documentclass[lettersize,journal]{IEEEtran}

\usepackage{amsmath,amsfonts}
\usepackage{algorithmic}
\usepackage{algorithm}
\usepackage{array}
\usepackage[caption=false,font=normalsize,labelfont=sf,textfont=sf]{subfig}
\usepackage{textcomp}
\usepackage{stfloats}
\usepackage{url}
\usepackage{verbatim}
\usepackage{graphicx}
\usepackage{cite}
\usepackage{afterpage}
\usepackage{diagbox}
\usepackage{pifont}  
\usepackage{romannum}
\usepackage{xcolor}
\usepackage{wrapfig}
\usepackage{graphicx}
\usepackage[colorlinks=true, linkcolor=blue, citecolor=blue, urlcolor=blue]{hyperref}

\begin{document}

\title{Performance Analysis of Underwater Optical Wireless Communication Using O-RIS and Fiber Optic Backhaul \\ \textit{(Extended version)}}

\author{Aboozar Heydaribeni and Hamzeh Beyranvand
\thanks{Aboozar Heydaribeni and Hamzeh Beyranvand are with the Department of Electrical Engineering,
Amirkabir University of Technology (Tehran Polytechnic), Tehran, Iran
(email: aboozar.hb98@aut.ac.ir; beyranvand@aut.ac.ir).}%
}

\IEEEtitleabstractindextext{%
\begin{abstract}
This Letter presents a novel hybrid underwater wireless optical communication (UWOC) system that integrates underwater optical access points (UOAPs) with a passive optical network (PON)-based fiber-optic backhaul to provide a resilient backbone. A hard switching mechanism is employed between direct and optical reconfigurable intelligent surface (O-RIS)-assisted links to ensure reliable connectivity. Unlike previous studies, the proposed system is evaluated under both active and multiple passive O-RIS configurations. To enhance reliability, the Selection Combining (SC) and Maximal Ratio Combining (MRC) schemes are applied. Analytical and simulation results demonstrate that optimal O-RIS placement significantly enhances system performance. However, in the linear regime, placing it too close to the receiver causes degradation due to increased path loss and beam jitter in an identical water type. Moreover, increasing the number of O-RIS elements within practical limits further improves overall system performance and enhances adaptability to variations in the underwater channel.
\end{abstract}
} 

\markboth{}%
{Heydaribeni \MakeLowercase{\textit{et al.}}: Performance Analysis of UWOC Using O-RIS and Fiber Optic Backhaul}

\maketitle
\IEEEdisplaynontitleabstractindextext
\IEEEpeerreviewmaketitle
\vspace{-1.5\baselineskip}

\section{Introduction}
\IEEEPARstart{U}{nderwater} wireless optical communication (UWOC) is recognized as a leading method for high-speed data transmission in aquatic environments. By utilizing light as the information carrier, it offers significant advantages over acoustic and RF systems, including higher bandwidth, lower latency, and enhanced security~\cite{kaushal2016underwater,elfikky2024underwater}. UWOC is crucial for underwater environmental monitoring, high-resolution imaging, and communication with devices such as robots, submarines, and sensors, with applications in marine biology, resource exploration, and environmental monitoring. It also plays a key role in coastal cities by managing marine resources, monitoring pollution, and observing ecosystems.
Optical reconfigurable intelligent surfaces (O-RISs), which can intelligently redirect optical waves, have recently emerged as a cost-effective and energy-efficient alternative to traditional relays in UWOC systems~\cite{liaskos2018new}. Furthermore, O-RIS can improve performance under line-of-sight (LOS) conditions and significantly enhance link reliability in non-line-of-sight (NLOS) channels by exploiting reflected paths to bypass obstacles and mitigate shadowing effects. A comprehensive comparison between IRS and relay architectures in terms of system performance was presented by H. Ajam \textit{et al.} \cite{10295484}, demonstrating that placing the relay midway between the transmitter (Tx) and receiver (Rx) minimized the outage probability. Collectively, these studies indicated that while relays typically excelled at high signal-to-noise ratios (SNRs), O-RISs exhibited superior performance at low SNRs due to reduced power consumption and hardware complexity.
{Table~\MakeUppercase{\romannumeral 1} summarizes recent key studies on UWOC, emphasizing the novelty of the proposed integrated methodology and the comprehensive evaluation of relevant parameters.
\begin{table}[tbp]
\centering
\setlength{\tabcolsep}{2.7pt}
\caption{\scriptsize Summary of Related Works, OP: Outage Probability, BER: Bit Error Rate, SOC: Speed Of Convergence, DO: Diversity Order, CC: Channel Capacity, SC: Selection Combining, MRC: Maximal Ratio Combining}
\vspace{-2mm}
\scriptsize
\begin{tabular}{|c|c|c|c|c|c|c|c|c|c|c|c|}
\hline
\diagbox{Aspect}{Ref.} & \cite{ata2024intelligent} & \cite{naik2022evaluation} & \cite{ata2023intelligent} & \cite{8606206} & \cite{wang2021performance} & \cite{zhang2024performance} & \cite{10413214}& \cite{10210651} & \cite{10413587}& \cite{10897965} & \begin{tabular}[c]{@{}c@{}}This\\ study\end{tabular} \\ \hline
\multicolumn{1}{|c|}{UWOC} & \checkmark & \checkmark & \checkmark & \checkmark & \ding{55} & \checkmark & \checkmark & \checkmark & \checkmark & \checkmark & \checkmark \\ \hline
\multicolumn{1}{|c|}{O-RIS} & \checkmark & \checkmark & \checkmark & \ding{55} & \checkmark & \checkmark & \checkmark & \checkmark & \checkmark & \checkmark & \checkmark \\ \hline
\multicolumn{1}{|c|}{OP} & \checkmark & \checkmark & \checkmark & \checkmark & \checkmark & \checkmark & \checkmark & \checkmark & \checkmark & \checkmark & \checkmark \\ \hline
\multicolumn{1}{|c|}{BER} & \ding{55} & \checkmark & \ding{55} & \checkmark & \checkmark & \ding{55} & \checkmark & \checkmark & \checkmark & \checkmark & \checkmark \\ \hline
\multicolumn{1}{|c|}{DO/SOC} & \ding{55} & \ding{55} & \ding{55} & \ding{55} & \ding{55} & \checkmark & \ding{55} & \ding{55} & \ding{55} & \ding{55} & \checkmark \\ \hline
\multicolumn{1}{|c|}{CC} & \ding{55} & \checkmark & \ding{55} & \checkmark & \ding{55} & \checkmark & \checkmark & \ding{55} & \checkmark & \checkmark & \checkmark \\ \hline
\multicolumn{1}{|c|}{\begin{tabular}[c]{@{}c@{}}Evaluation over \\ Different Link Lengths \end{tabular}} & \checkmark & \ding{55} & \checkmark & \ding{55} & \ding{55} & \ding{55} & \ding{55} & \ding{55} & \checkmark & \checkmark & \checkmark \\ \hline
\multicolumn{1}{|c|}{\begin{tabular}[c]{@{}c@{}} O-RIS Placement vs. \\ System Performance\end{tabular}} & \ding{55} & \ding{55} & \ding{55} & \ding{55} & \ding{55} & \ding{55} & \ding{55} & \ding{55} & \ding{55} & \ding{55} & \checkmark \\ \hline
\multicolumn{1}{|c|}{\begin{tabular}[c]{@{}c@{}}Comparison of Active \\ and Passive O-RIS\end{tabular}} & \ding{55} & \ding{55} & \ding{55} & \ding{55} & \ding{55} & \ding{55} & \ding{55} & \ding{55} & \ding{55} & \ding{55} & \checkmark \\ \hline
\multicolumn{1}{|c|}{\begin{tabular}[c]{@{}c@{}}Comparison of SC \\ and MRC Techniques\end{tabular}} & \ding{55} & \ding{55} & \ding{55} & \ding{55} & \ding{55} & \ding{55} & \ding{55} & \ding{55} & \ding{55} & \ding{55} & \checkmark \\ \hline
\multicolumn{1}{|c|}{Fiber optic Backhaul} & \ding{55} & \ding{55} & \ding{55} & \ding{55} & \ding{55} & \ding{55} & \ding{55} & \ding{55} & \ding{55} & \ding{55} & \checkmark \\ \hline
\end{tabular}
\end{table}

As the performance of O-RIS-assisted long-range UWOC systems strongly depends on accurate characterization of optical turbulence for this special medium, the Gamma–Gamma (G–G) distribution, characterized by the Oceanic Turbulence Optical Power Spectrum (OTOPS) and employed in this study as well as in~\cite{10897965}, provides higher accuracy and analytical capability compared to recently proposed models such as the $\lambda$–$\kappa$–$\mu$ and double $\lambda$–$\kappa$–$\mu$ distributions reported in~\cite{11105415}.

The motivation for this work arises from the limitations of long-range UWOC links in offshore scenarios, where optical attenuation, scattering, and turbulence severely degrade link reliability and throughput. While standalone UWOC or fiber systems each face scalability and deployment constraints, hybrid solutions such as hybrid RF–UWOC or acoustic–UWOC often trade data rate for robustness. In contrast, integrating an O-RIS-assisted UWOC front-end with a fiber backhaul offers a balanced and practical solution: the O-RIS improves optical link stability and spatial flexibility, whereas the fiber ensures high-capacity and reliable core connectivity.

Therefore, the key contribution is the proposal of a hybrid UWOC architecture integrating intelligent underwater optical access points (UOAPs), a PON-based fiber backhaul, and O-RISs. PON combined with DWDM structures, as was shown by E.~E.~Elsayed \cite{elsayed2024performance}, significantly improved bandwidth utilization and maintained stable high-capacity links, thereby forming a robust and scalable backbone.
O-RISs, deployed in active or multiple passive configurations, reconfigure optical wave propagation to extend coverage and mitigate misalignment or obstacles in both LOS and NLOS channels. 

A hard-switching mechanism enables fast and deterministic transitions between the direct link and the O-RIS-assisted link to maintain seamless connectivity, while diversity combining techniques such as selection combining (SC) and maximal ratio combining (MRC) improve link robustness and effectively mitigate channel fading. 
By integrating these key features, the proposed system delivers a unified and practical solution that bridges the gap between
high-capacity and highly reliable UWOC, marking a notable step toward operational long-range underwater networks.

\vspace{-4mm}
\section{System Model}
As illustrated in Fig. 1, the proposed architecture employs a fiber-optic backhaul network, where the Optical Line Terminal (OLT) is connected to a Passive Optical Splitter (POS), which passively divides optical signals into multiple downstream paths and combines upstream signals toward the OLT, while Wavelength Division Multiplexing (WDM) in the network is used for duplexing (1358/1286 nm for DL/UL). Each splitter branch connects via a fiber-optic path to an Optical Network Unit (ONU), which links to the UOAP as the backhaul gateway. The UOAP, equipped with five strategically placed transceivers (left, right, front, back, and top), ensures full spatial coverage.
Communication between the UOAP and the ONU follows a TDM-based data allocation strategy, using Dynamic Bandwidth Allocation (DBA) for optimized bandwidth and Forward Error Correction (FEC) to reduce transmission errors. Integrating UWOC, O-RIS, and fiber-optic backhaul, the proposed architecture mitigates obstructions, turbulence, and attenuation, enhancing range and data transmission.
\begin{figure*}
    \begin{minipage}{\textwidth}
        \centering
        \includegraphics[width=\linewidth, keepaspectratio]{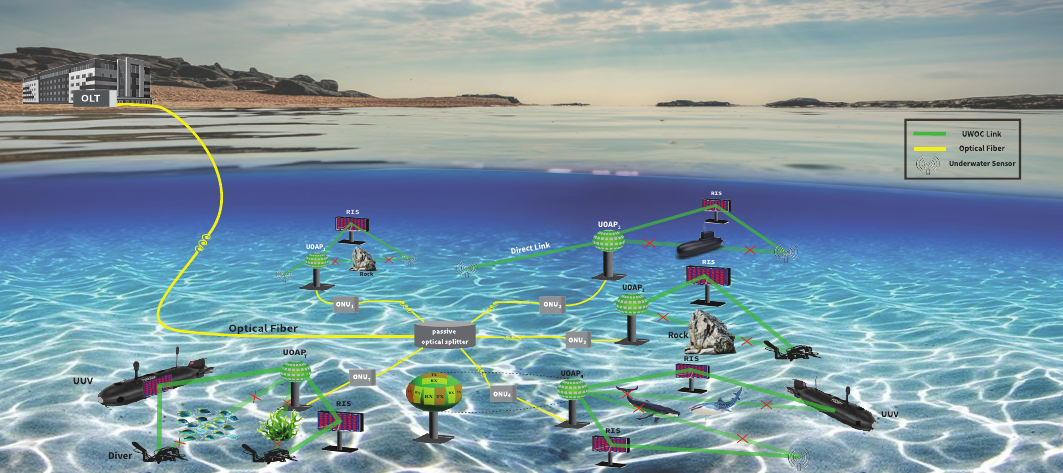}
        \captionsetup{justification=centering}
        \caption{\small Proposed research system model illustrating the unmanned underwater vehicles (UUV), transmitter (Tx), and receiver (Rx).}
        \label{fig:system-model}
    \end{minipage}
\end{figure*} 

Throughout this letter, the O-RIS is considered to operate in the linear regime \cite{10295484}. A beam expander is assumed on the transmitter side to ensure a uniform power distribution over each O-RIS element (Fig. 3 in \cite{wang2021performance}). On the receiver side, a lens with an aperture larger than the beam footprint is used to concentrate the received power and maximize the beamforming gain \cite{wang2021performance,10295484}.
The considered UWOC system incorporates an O-RIS with \( N \) reflective elements, and the received signal \( r \) is expressed as a synthesis of multiple components, given by: $r = \sum_{m=1}^{N} H_m x + w,( H_m = h_{\alpha_{s \cdot r_m}} \cdot h_{l_{s \cdot r_m}} \cdot h_{\alpha_{r_m \cdot d}} \cdot h_{l_{r_m \cdot d}} \cdot h_{P_m} \cdot \rho_m \cdot e^{i \theta_m})$;
where \( H_m \) represents the composite channel model for the \( m \)-th element of the O-RIS, incorporating channel effects for the two-hop UWOC link. The attenuation (\( h_l \)) in the O-RIS--assisted NLOS link follows the Beer–Lambert law, expressed as $h_{l_m}=e^{-c.L_{s.d}}$, while the LOS link is modeled as $h_l = e^{-c.L_D} $~\cite{elfikky2024underwater}. Water surface fluctuations cause O-RIS jitter by altering reflection angles, leading to propagation mismatch. Additionally, fish movements and water currents introduce scattering and diffraction, resulting in beam misalignment (\( h_p \)) that is modeled using the Rayleigh distribution, as detailed in Eqs.~(6) and (7) of~\cite{shetty2023performance} for the direct link, and in Eqs.~(5) and (12) of ~\cite{wang2021performance} for the O-RIS--assisted link. Since the O-RIS size and element spacing are much smaller than the link distance, all elements experience identical attenuation and pointing error. Thus, \( h_{l_m} = h_l \) and \( h_{p_m} = h_p \), \( \forall m = 1, \dots, N \).

The direct channel turbulence fading \( (h_{\alpha_D}) \) is modeled using a Gamma distribution to represent single-path turbulence. In contrast, the O-RIS--assisted cascaded channel fading coefficient \( h_{\alpha_\text{O-RIS},m} = h_{\alpha_{s.r_m}} h_{\alpha_{r_m.d}}\), which captures independent turbulence effects over the source–O-RIS and O-RIS–destination links, follows a G–G distribution, as it is the product of two independent Gamma-distributed random variables (RVs) (see Appendix~D). Moreover, for comparison and to enhance the comprehensiveness of the study, the mixture Exponential-Generalized Gamma (mEGG) model is also evaluated based on~\cite{10413214}. According to the approximation of the sum of correlated G–G RVs, if {\small\( X_n \sim \Gamma\Gamma(A, B, \bar{\Omega}) \)} and the correlation coefficient between them is \( \rho_{_{nm}} \), the sum can be approximated as {\small\( Y \sim \Gamma\Gamma(c\alpha, c\beta, c\Omega) \)}, where {\small\( c = N \left(1 + \frac{2}{N} \sum_{n=1}^{N} \sum_{m=n+1}^{N} \rho_{_{nm}} \right)^{-1} \)}. 
Assuming the RVs are i.i.d. (i.e., $\rho_{_{nm}} = 0$), we have $c = N$, and therefore the shape parameters of the equivalent distribution follow $A = N.\alpha$ and $B = N.\beta$ (see Appendix F and  ~\cite[Appendix A]{8438298})~\cite{zhang2024performance}. Moreover, for the mEGG distribution, as shown in~\cite{10413214}, the parameter $N$ has a linear relationship with the shape parameter, given by
$m = N.\frac{(E[|H_m|])^2}{\text{Var}(|H_m|)}$, thereby directly affecting the SNR distribution and the overall system performance. The environmental parameters of UWOC for the G–G model (which differ significantly from those used in FSO analysis) as well as the mEGG model, accurately reflect the scintillation index ($\sigma_I^2$) based on~\cite{fu2018performance} and ~\cite{8606206}, respectively.
Also environmental noise effects (\(w\)) are modeled as AWGN~\cite{ata2024intelligent}, with zero mean and variance \(\sigma_n^2\), i.e., \(w \sim \mathcal{N}(0, \sigma_n^2)\). 
The instantaneous SNRs of the passive and active O-RIS--assisted systems, with $\bar{\gamma} = \frac{P_t \cdot h_l^2}{\sigma_n^2}$, are calculated as follows:

\vspace{-4mm}
{ \scriptsize
\begin{equation}
\centering
\scalebox{0.95}[1.1]{%
$
\gamma_{_\text{Pas}} = \bar{\gamma}. h_p^{2} . \left( {\displaystyle\sum_{m=1}^{N}} \rho_{m}.e^{j\theta_{m}}.h_{\alpha,m} \right)^{2}, \hspace{2mm}
\gamma_{_\text{Act}} = \dfrac{\bar{\gamma}.h_p^{2}. \left( {\displaystyle\sum_{m=1}^{N}} G_m.e^{j\theta_{m}}.h_{\alpha,m} \right)^{2}}{1 + N. P_{\text{noise\_Act.\_element}}}
$
}
\end{equation}}
where $\rho$, $c$, $\theta_m$, $G_m$, $\mathrm{i}$, $x$, $P_t$, $D$, $s$ and $d$ denote the reflection coefficient, attenuation coefficient, phase shift, active gain, imaginary unit, transmitted signal, average transmit power, direct, source and destination, respectively. Finally, the probability density function (PDF) of the instantaneous SNR is derived in detail. The O-RIS results are provided in Appendix D, and a similar approach is also adopted for the direct link.
\section{System Performance Analysis}
The performance of the proposed system is evaluated using key closed-form metrics, including Outage Probability (OP), Bit Error Rate (BER), Diversity Order (DO), Speed of Convergence (SOC), and Channel Capacity (CC) for the O-RIS. Among these, OP is a primary indicator of link reliability under underwater turbulence and is examined for two representative turbulence models. Furthermore, the total system capacity is analyzed to capture the impact of dynamic UOAP configurations and integration with the fiber-optic backhaul.

\textbf{A. OP:} OP quantifies the likelihood of communication failure and serves as a key metric for ensuring link reliability. An outage event occurs when the instantaneous signal-to-noise ratio (SNR), ($\gamma$), falls below a predefined threshold ($\gamma_{\mathrm{th}}$). This Letter evaluates the OP of a direct link, modeled using the Gamma turbulence model, and an O-RIS--assisted link, analyzed under the mEGG and G–G models. The OP for the mEGG model is obtained from Eq.~(22) in \cite{10413214}, whereas for the Gamma and G–G models, it is derived in Eqs.~(2) and~(3) of this letter, based on Appendix~D. By default, the UOAP system is designed such that the direct and O-RIS--assisted links operate independently, with only one link active for each Tx/Rx pair at any given time. Typically, the direct link is utilized, while the O-RIS--assisted link remains inactive, serving as a fallback upon blockage. However, the system supports dynamic reconfiguration, enabling simultaneous activation of both links to enhance QoS. Assuming i.i.d. fading and full coherence between the received signals, diversity techniques such as SC and strategies closely resembling MRC can be applied. The OPs for both combining schemes are derived as follows: {\small $\gamma_{SC}=max{\left(\gamma_D,\gamma_{O-RIS}\right)}; \gamma_{\mathrm{MRC}}=\gamma_D+\gamma_{{\tiny{O-RIS}}}; OP_{\mathrm{SC}}=Pr\left(\gamma_{\mathrm{SC}}\le\gamma_{\mathrm{th}}\right)=F_{\gamma_D}\left(\gamma_{\mathrm{th}}\right)\cdot F_{\gamma_{\mathrm{O-RIS}}}\left(\gamma_{\mathrm{th}}\right)$}.

\noindent {\scriptsize
\begin{equation}
\makebox[\linewidth][l]{%
  \scalebox{0.73}[1.1]{%
    \begin{minipage}{\linewidth}
    \setlength{\abovedisplayskip}{0pt}
    \setlength{\belowdisplayskip}{0pt}
    \begin{equation*}
    \begin{aligned}
    OP_D(\bar{\gamma}) = 
    &\left( \frac{\alpha}{A_0 \sqrt{\bar{\gamma}}} \right)^{\xi^2} 
    \times \frac{\xi^2 \cdot 2^{\alpha - \xi^2 - 2}}{\Gamma(\alpha) \cdot \sqrt{\pi}} 
    \times \gamma_{th}^{\frac{\xi^2}{2}} &\times G_{3,5}^{4,1} \left(
    \frac{\alpha^2 \gamma_{th}}{4 A_0^2 \bar{\gamma}} \,\Bigg|\, 
    \begin{array}{l}
    1 - \frac{\xi^2}{2},\ \frac{1}{2},\ 1 \\
    \frac{\alpha - \xi^2}{2},\ \frac{\alpha - \xi^2 + 1}{2},\ 0,\ \frac{1}{2},\ -\frac{\xi^2}{2}
    \end{array}
    \right)
    \end{aligned}
    \end{equation*}
    \end{minipage}
  }%
} \label{formula:2}
\end{equation}
\noindent
\begin{equation}
\makebox[\linewidth][l]{%
  \scalebox{0.72}[1.1]{%
    \begin{minipage}{\linewidth}
    \setlength{\abovedisplayskip}{0pt}
    \setlength{\belowdisplayskip}{0pt}
    \begin{equation*}
    \begin{aligned}
     OP_{O-RIS}(\bar{\gamma}) = \frac{2^{A+B-5} AB \xi^2}{\pi N \Gamma(A) \Gamma(B) A_0} 
    \left( \frac{\gamma_{th}}{\bar{\gamma}} \right)^{\frac{1}{2}} 
    \times G_{3,7}^{6,1} \left(
    \frac{A^2 B^2 \gamma_{th}}{16 N^2 A_0^2 \bar{\gamma}} \Bigg| 
    \begin{array}{l}
    \frac{1}{2}, \frac{\xi^2}{2}, \frac{\xi^2 + 1}{2} \\
    \frac{\xi^2 - 1}{2}, \frac{\xi^2}{2}, \frac{A - 1}{2}, \frac{B - 1}{2}, \frac{A }{2}, \frac{B }{2}, -\frac{1}{2}
    \end{array}
    \right)
    \end{aligned}
    \end{equation*}
    \end{minipage}
  }%
} \label{formula:3}
\end{equation} 
\vspace{-5mm}
\begin{equation}
\makebox[\linewidth][l]{%
  \scalebox{0.585}[1.1]{%
    \begin{minipage}{\linewidth}
    \[
    \begin{aligned}
     OP&_{MRC}(\bar{\gamma}) = \int_{0}^{\gamma_{th}}  F_{\gamma_{O-RIS}}\left(\gamma_{th}-\gamma_D\right) f_{\gamma_D}\left(\gamma_D\right) d\gamma_D = \int_{0}^{\gamma_{th}} \frac{2^{A+B-5} AB \xi_R^2}{\pi N \Gamma(A) \Gamma(B) A_{0,R}} 
    \left(\frac{\gamma_{th}-\gamma_D}{\bar{\gamma}}\right)^{\frac{1}{2}} \frac{\xi_D^2}{2 \Gamma(\alpha_D)} \left(\frac{\alpha_D}{A_{0,D} \sqrt{\bar{\gamma}}}\right)^{\xi_D^2} \\
    &  \gamma_D^{\frac{\xi_D^2}{2} - 1} \times G_{3,7}^{6,1}\left(\frac{A^2 B^2 \left(\gamma_{th}-\gamma_D\right) }{16 N^2 A_{0,R}^2 \bar{\gamma}} \middle| 
    \begin{array}{l}
    \frac{1}{2}, \frac{\xi_R^2}{2}, \frac{\xi_R^2 + 1}{2} \\
    \frac{\xi_R^2 - 1}{2}, \frac{\xi_R^2}{2}, \frac{A-1}{2}, \frac{B-1}{2}, \frac{A}{2}, \frac{B}{2}, -\frac{1}{2}
    \end{array}
    \right) \times G_{1,2}^{2,0}\left(\frac{\alpha_D}{A_{0,D} \sqrt{\bar{\gamma}}} \gamma_D^{1/2} \middle|
    \begin{array}{l}
    1 \\
    0, \alpha_D - \xi_D^2
    \end{array}
    \right)
    d\gamma_D
    \end{aligned}
    \] \label{formula:4}
    \end{minipage}
  }
}
\end{equation} }
\noindent\textbf{B. BER:} BER is a key metric for evaluating signal quality and transmission accuracy, directly influencing overall system performance. In the proposed system, various modulation schemes such as BPSK, BFSK, Q-PSK, and 16-QAM are employed. The BER for the proposed system is computed using the following expression: {\small $\displaystyle \text{BER} =\frac{\delta}{{2}\Gamma({p})} \sum_{k=1}^{\mathcal{K}}[{q}_{k}^{p}\int_{{0}}^{\infty}{{e}^{-{q}_{k}\gamma}\gamma^{{p}-{1}}{F}_\gamma(\gamma)}{d}\gamma]\ $}as derived in \cite{ramavath2020co}. The parameters $\delta$, $p$, $q_k$, and $\mathcal{K}$ depend on the modulation type, as detailed in Table 1 of \cite{ramavath2020co}. This formulation applies to both direct and O-RIS--assisted scenarios (see Appendix~D); the BER is calculated as follows:

{
\scriptsize 
\begin{equation}
\makebox[\linewidth][l]{%
  \scalebox{0.72}[1.1]{%
$\displaystyle \textit{BER}_D = 
\frac{\delta}{2\Gamma(p)} \sum\limits_{k=1}^{\mathcal{K}}  
 \frac{\zeta^2 \cdot 2^{\alpha - \zeta^2 - 2}}{\sqrt{\pi} \Gamma(\alpha)}
\left( \frac{\alpha}{A_0 \sqrt{q_k \bar{\gamma}}} \right)^{\zeta^2} \cdot
G_{4,5}^{4,2} \left(
\frac{\alpha^2}{4 A_0^2 q_k \bar{\gamma}} \left|
\begin{array}{c}
1 - p - \frac{\zeta^2}{2},\ 1 - \frac{\zeta^2}{2},\ \frac{1}{2},\ 1 \\
\frac{\alpha - \zeta^2}{2},\ \frac{\alpha - \zeta^2 + 1}{2},\ 0,\ \frac{1}{2},\ -\frac{\zeta^2}{2}
\end{array}
\right.
\right)$}}
\end{equation}
\begin{equation}
\makebox[\linewidth][l]{%
  \scalebox{0.67}[1.1]{%
$\displaystyle
\textit{BER}_{\mathrm{O\text{-}RIS}} =
\frac{\delta}{2 \Gamma(p)} 
\sum_{k=1}^{\mathcal{K}} 
\frac{2^{A + B - 5} \cdot AB \cdot \zeta^2}{\pi N A_0 \Gamma(A) \Gamma(B) \sqrt{q_k \bar{\gamma}}}
\cdot
G_{4,7}^{6,2} \left(
\frac{A^2 B^2}{16 q_k N^2 A_0^2 \bar{\gamma}}
\left|
\begin{array}{c}
\frac{1}{2},\, \frac{1}{2} - p,\, \frac{\zeta^2}{2},\, \frac{\zeta^2 + 1}{2} \\
\frac{\xi^2 - 1}{2},\, \frac{\xi^2}{2},\, \frac{A - 1}{2},\, \frac{B - 1}{2},\, \frac{A}{2},\, \frac{B}{2},\, -\frac{1}{2}
\end{array}
\right.
\right)$ }}
\end{equation}
}

\noindent\textbf{C. DO and SOC:} DO is a key metric for mitigating multipath attenuation and channel fading by utilizing multiple independent transmission paths, thereby enhancing link reliability and signal quality. The Asymptotic Diversity Order (ADO) characterizes performance in the high-SNR regime, whereas SOC quantifies how rapidly the system adapts to environmental changes and achieves steady performance, which is critical for real-time communication. To complete the system analysis, the definitions of DO, ADO, and SOC are given as follows (as presented analogously in Appendix~D):

{
\scriptsize
\begin{equation}
\makebox[\linewidth][c]{%
  \scalebox{0.85}[1.1]{%
    $\displaystyle
    DO(\bar{\gamma}) = -\frac{\partial \ln \mathrm{OP}_{\text{O-RIS}}(\bar{\gamma})}{\partial \ln \bar{\gamma}}, \hspace{3pt}
    ADO = \frac{1}{2} \times \min(A,\xi^2,B), \hspace{3pt}
    SOC(\bar{\gamma}) = \frac{1}{ADO} \frac{\partial DO(\bar{\gamma})}{\partial \ln \bar{\gamma}}.
    $%
  }%
}
\label{eq:DO_SOC}
\end{equation}
}
\noindent\textbf{D. CC:} CC represents the maximum achievable data rate ensuring signal quality and reliability. For both direct and O-RIS--assisted links, the ergodic capacity is obtained in closed form using standard integral transformations (see Appendices~D and~E). In this model, the total capacity is limited by the minimum of the UOAP link capacity and the PON backhaul. In the absence of PON constraints, the expression reduces to {\small$C_{\text{total}}(\bar{\gamma}) = T \cdot C_{\text{UOAP}}(\bar{\gamma})$}. This corresponds to a bandwidth of $100$~MHz, an aggregate spectral efficiency of $150$~bps/Hz, and an approximate data rate of $15$~Gbps ({\small$C_{\max}$}) per UOAP over standard {\small ITU-T G.652.D} single-mode fiber to the POS, where $u$ and $v$ denote the numbers of direct and O-RIS-assisted links in each UOAP, respectively, with the constraint {\small $u+v \le 5$}.} 
Furthermore, as shown in Eq.~(11), the throughput per UOAP is defined to measure the efficiency of data transmission in underwater networks.

{\scriptsize
\begin{equation}
\makebox[\linewidth][l]{%
  \scalebox{0.72}[1.1]{%
    \begin{minipage}{\linewidth}
    \setlength{\abovedisplayskip}{0pt}
    \setlength{\belowdisplayskip}{0pt}
    \begin{equation*}
    \begin{aligned}
    \hspace{0.5cm}C_D(\bar{\gamma}) = \left( \frac{\alpha}{A_0 \sqrt{\tau \bar{\gamma}}} \right)^{\zeta^2}
    \frac{\zeta^2 2^{(\alpha - \zeta^2 - 2)}}{\sqrt{\pi} \Gamma(\alpha) \ln(2)} 
   \times G_{4,6}^{6,1} \left( 
   \frac{\alpha^2}{4 A_0^2 \tau \bar{\gamma}} 
   \Bigg| 
   \begin{array}{c}
   \frac{-\zeta^2}{2}, 1 - \frac{\zeta^2}{2}, \frac{1}{2}, 1 \\
   \frac{\alpha - \zeta^2}{2}, \frac{\alpha - \zeta^2 + 1}{2}, 0, \frac{1}{2}, -\frac{\zeta^2}{2}, -\frac{\zeta^2}{2}
   \end{array}
    \right)
    \end{aligned}
    \end{equation*} \label{formula:9}
    \end{minipage}
  }%
}
\end{equation}
\noindent 
\begin{equation}
\makebox[\linewidth][l]{%
  \scalebox{0.695}[1.1]{%
    \begin{minipage}{\linewidth}
    \setlength{\abovedisplayskip}{0pt}
    \setlength{\belowdisplayskip}{0pt}
    \begin{equation*}
    \begin{aligned}
C_{O-RIS}(\bar{\gamma}) = \frac{2^{A + B - 5} AB \zeta^2}{ \pi N \ln(2) \Gamma(A) \Gamma(B) A_0 \sqrt{\tau \bar{\gamma}}} \times G_{4,8}^{8,1} \left( 
\frac{A^2 B^2}{16 N^2 A_0^2 \tau \bar{\gamma}} 
\Bigg| 
\begin{array}{c}
\frac{-1}{2}, \frac{1}{2}, \frac{\zeta^2}{2}, \frac{\zeta^2 + 1}{2} \\
\frac{\xi^2 - 1}{2}, \frac{\xi^2}{2}, \frac{A - 1}{2}, \frac{B - 1}{2}, \frac{A }{2}, \frac{B }{2}, -\frac{1}{2}, -\frac{1}{2}
\end{array}
\right)
    \end{aligned}
    \end{equation*}
    \end{minipage}
  }%
} \label{formula:10}
\end{equation}
\vspace{-3mm}
\begin{equation}
\makebox[\linewidth][c]{%
    \scalebox{0.82}[1.1]{%
    $\begin{aligned}
    C_{UOAP}(\bar{\gamma}) = \sum_{i=1}^{u} C_{D_i}&(\bar{\gamma}) + \sum_{j=1}^{v} C_{O\text{-}RIS_j}(\bar{\gamma}) \, ; \hspace{2mm}
    C_{POS\text{-}UOAP}(\bar{\gamma}) = \frac{\eta \cdot C_{PON(OLT\text{-}POS)}}{T} ; \\[-1mm]
    &
    C_{Total}(\bar{\gamma}) = \min\left\{T \cdot C_{UOAP}(\bar{\gamma}) , \ C_{POS\text{-}UOAP}(\bar{\gamma})\right\}
    \end{aligned}$ 
  }
} \label{formula:11}
\end{equation}
\begin{equation}
\makebox[\linewidth][c]{%
    \scalebox{0.8}[1.1]{%
\(
\centering
\begin{aligned}
TH&_{\mathrm{real}}(i,T) = 
\min\Big(D_i,\min\Big(\frac{D_i}{\sum_{j=1}^{T}D_j}(\eta.R_{\mathrm{upstream}}),
\min(C_{\max},\frac{\eta.R_{\mathrm{upstream}}}{T})\Big)\Big) \times \Big(1- \\[-1mm] &\min(0.25,0.005(\max(\frac{\sum_{k=1}^{T}D_k}{\eta.R_{\mathrm{upstream}}},1))^{2.5})\Big),\hspace{1mm} D_i = \frac{1/i^\alpha}{\sum_{j=1}^{T} 1/j^\alpha} \times (1.6 \times R_{\mathrm{upstream}})
\end{aligned}
\)
}}  
\end{equation}
} 
\vspace{-7mm}
\section{ Quantitative Analysis and Interpretation} 
The parameters considered in this study are presented in Table \MakeUppercase{\romannumeral2}.  Applying O-RIS in UWOC faces challenges such as pressure, noise, and energy limits. To address these, this study employs a passive O-RIS with fixed phase and coefficient ($\rho$), while optimizing its angles, signal intensity, and placement.

\begin{figure*}[t]
\centering
\hspace{-2mm}
\includegraphics[width=0.25\textwidth, trim={9pt 0 20pt 20pt}, clip]{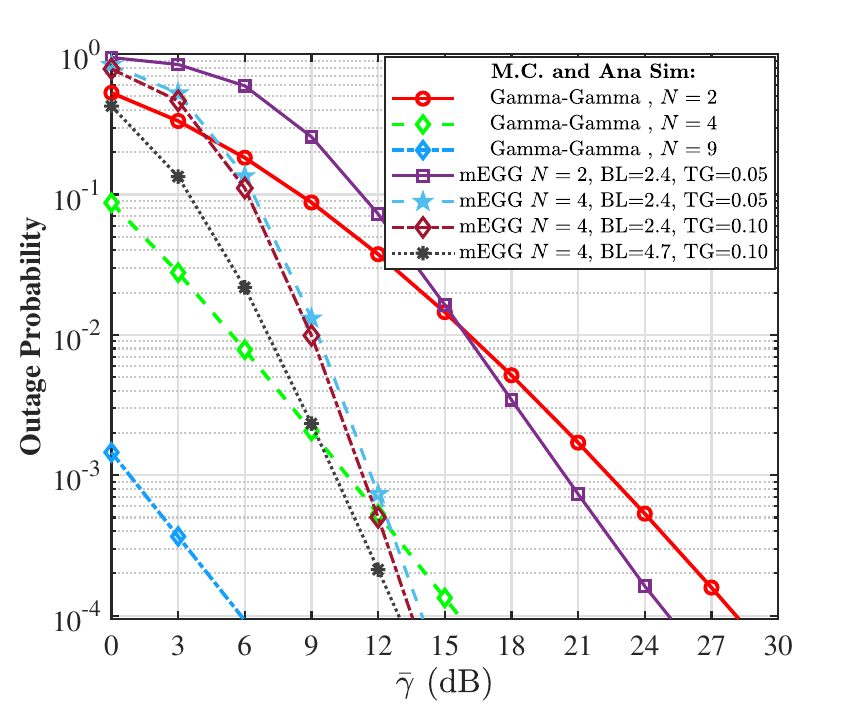}%
\includegraphics[width=0.242\textwidth, keepaspectratio, trim={10pt 0 30pt 15pt}, clip]{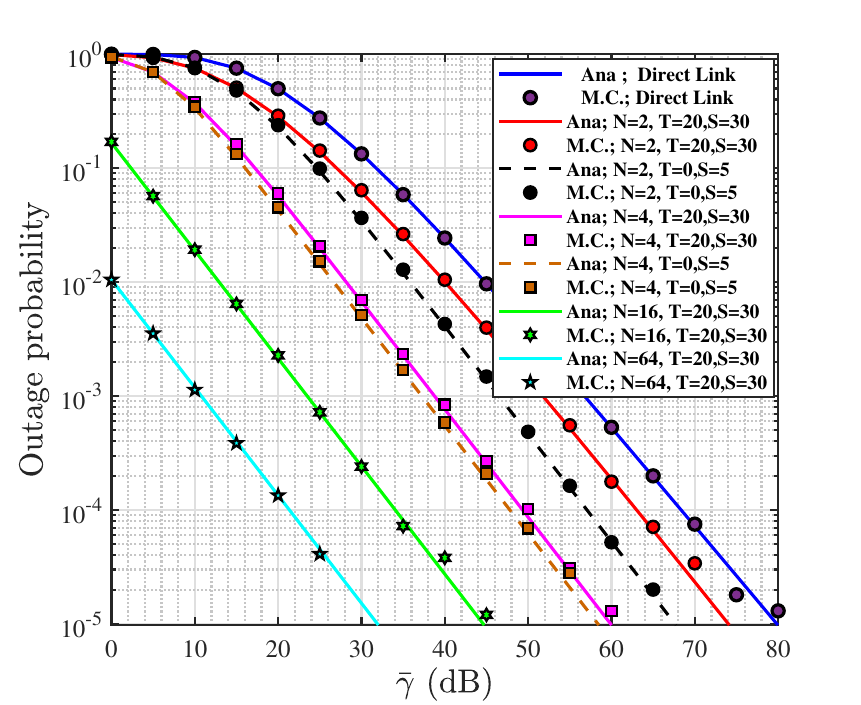}
\hspace{0mm}%
\includegraphics[width=0.25\textwidth, trim={5pt 0 22pt 18pt}, clip]{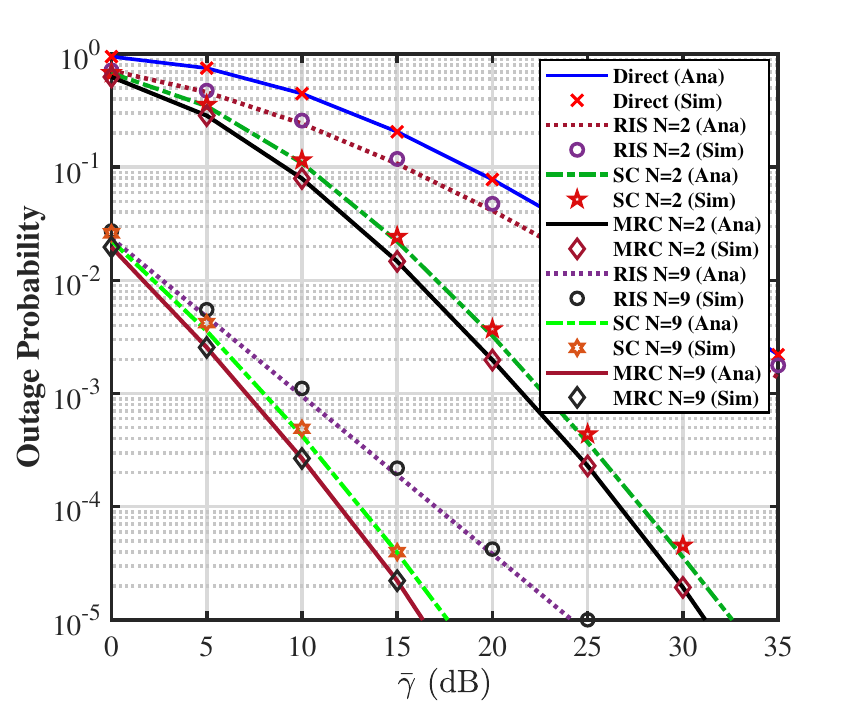}%
\includegraphics[width=0.242\textwidth, trim={15pt 0 25pt 18pt}, clip]{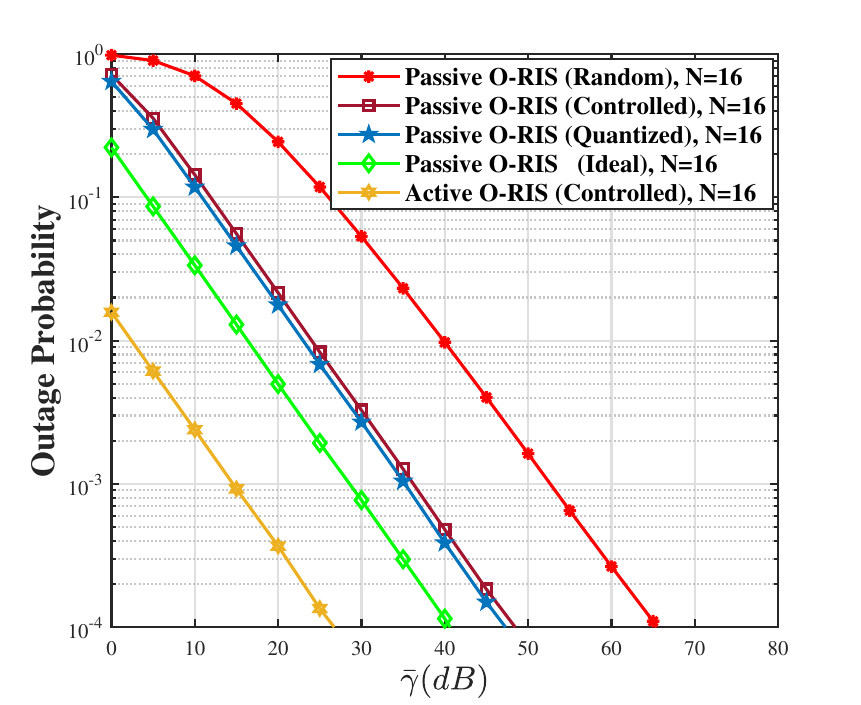}\\[-5pt]
\scriptsize
\makebox[0.251\textwidth][c]{\hspace{2mm}\textbf{(a)}}%
\makebox[0.251\textwidth][c]{\hspace{2mm}\textbf{(b)}}%
\makebox[0.251\textwidth][c]{\hspace{0mm}\textbf{(c)}}%
\makebox[0.251\textwidth][c]{\hspace{0mm}\textbf{(d)}}\\[-0.5pt]
\hspace{-1mm}
\includegraphics[width=0.245\textwidth, trim={15pt 0 25pt 20pt}, clip]{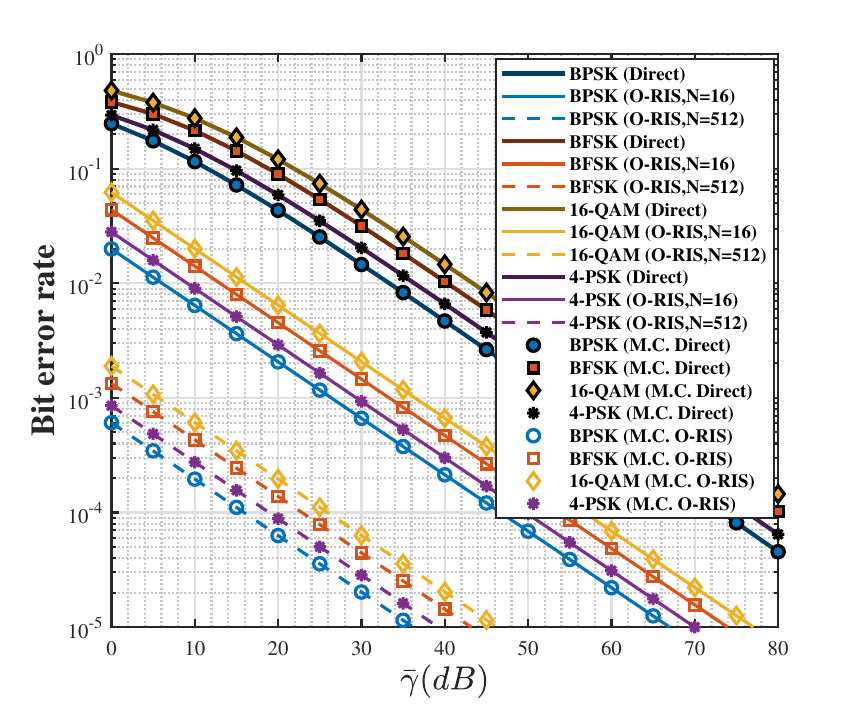}%
\hspace{1mm}%
\includegraphics[width=0.245\textwidth, trim={15pt 0 25pt 20pt}, clip]{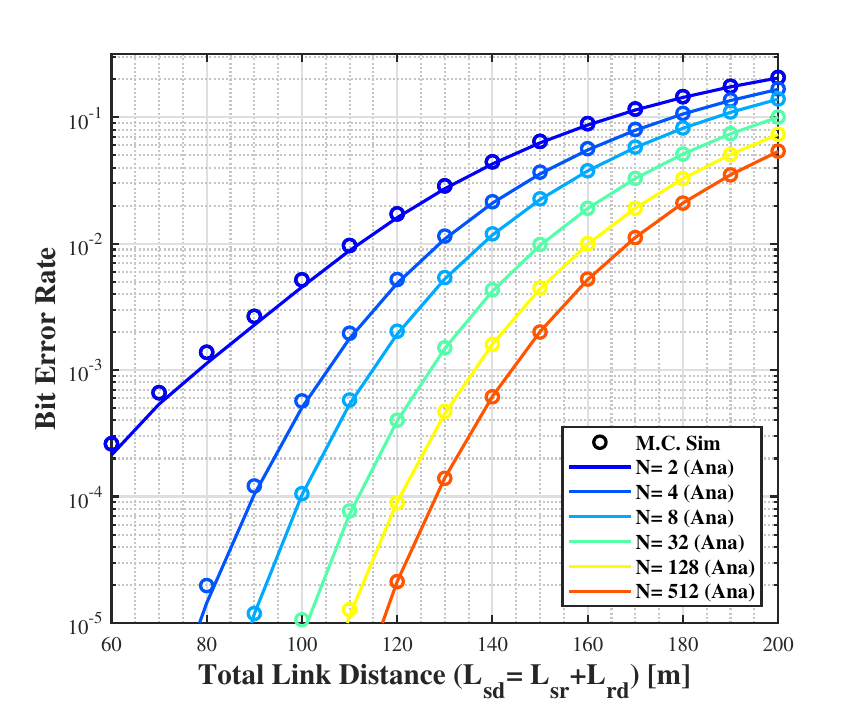}%
\hspace{0mm}%
\includegraphics[width=0.245\textwidth, trim={15pt 0 25pt 20pt}, clip]{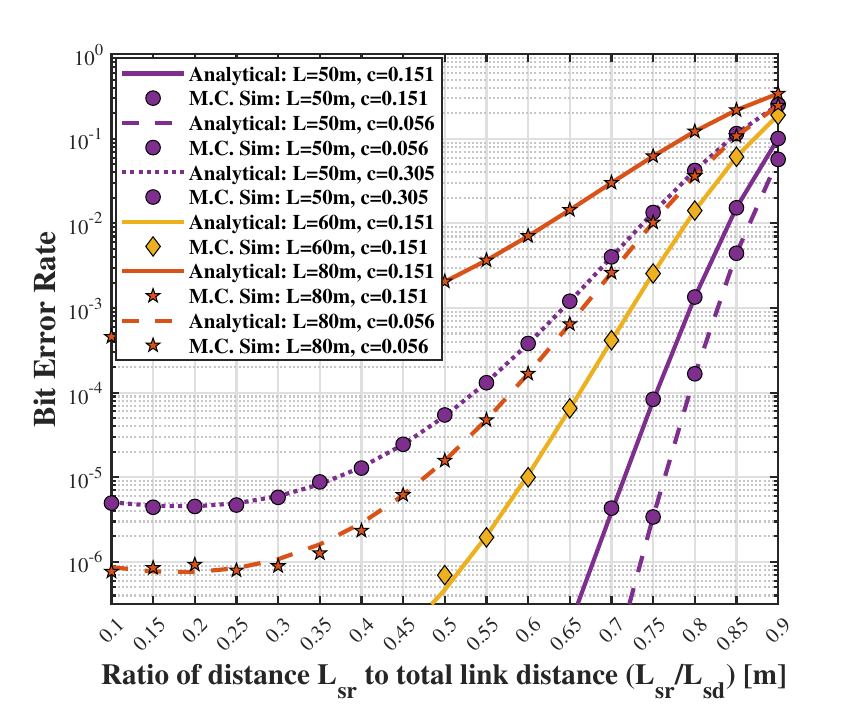}%
\hspace{0mm}%
\includegraphics[width=0.245\textwidth, trim={10pt 0 25pt 20pt}, clip]{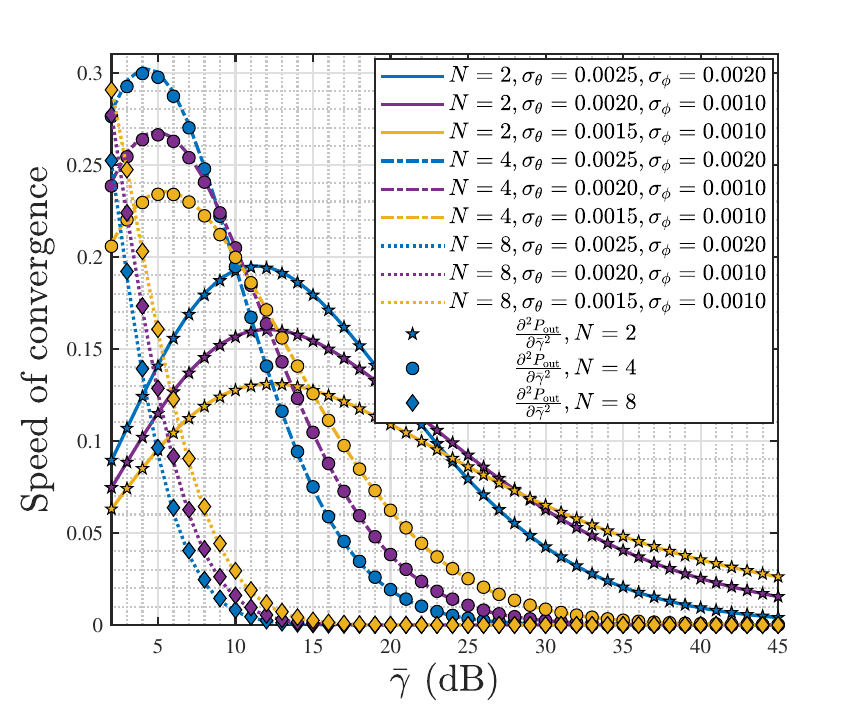}\\[-4pt]
\scriptsize
\makebox[0.251\textwidth][c]{\hspace{2mm}\textbf{(e)}}%
\makebox[0.251\textwidth][c]{\hspace{2mm}\textbf{(f)}}%
\makebox[0.251\textwidth][c]{\hspace{0mm}\textbf{(g)}}%
\makebox[0.251\textwidth][c]{\hspace{0mm}\textbf{(h)}}\\[-3.5mm]
\caption{(a) OP vs. average SNR ($\bar{\gamma}$) for $L_{sr}=L_{rd}=1\ \mathrm{m}$ under mEGG and G–G turbulence models. (b) OP vs. $\bar{\gamma}$ for various temperature and salinity levels. (c) OP vs. $\bar{\gamma}$ for $L_{sr}=L_{rd}=40\ \mathrm{m}$ and $\gamma_{th}=5$ dB under SC and MRC techniques. (d) OP comparison vs. $\bar{\gamma}$ for \( L_{sr} = L_{rd} = 50\ \mathrm{m} \) with various O-RIS types. (e) BER vs. $\bar{\gamma}$ for different modulation types. (f) BER of BPSK vs. {\small$L_{sd}$}, with $L_{sr}=L_{rd}$ and $\bar{\gamma}=25~\mathrm{dB}$. (g) BER of BPSK vs. {\small$(\frac{L_{sr}}{L_{sd}})$} for various water types, with $P_t = 20~\mathrm{dB}$ and $N_{\mathrm{O\text{-}RIS}} = 16$. (h)  SOC vs. $\bar{\gamma}$ for under different {\scriptsize$N_{O\text{-}RIS}$} and jitter conditions, with \( L_{sr} = L_{rd} = 60\ \mathrm{m} \).}
\label{fig:super_compact}
\end{figure*}
In Fig. 2(a), the statistical models, mEGG and G–G, are compared to evaluate the performance of the system under underwater optical turbulence. The mEGG model, due to its composite structure, offers a more accurate representation of the environment, particularly at high SNRs. However, it lacks a closed-form relationship between its parameters $(\omega, \lambda, a, b,c)$ and the link distance. These parameters are typically derived from experimental scintillation index ($\sigma_I^2$) data and are numerically estimated using iterative methods such as the expectation maximization (EM) algorithm. For instance, in~\cite{8606206}, they were estimated for a 1-meter link under laboratory conditions. Applying the model to different link distances or environmental settings requires new measurements and a similarly complex and computationally demanding parameter estimation process.
In this letter, we emphasize rigorous analytical modeling and employ the G–G model to enable accurate and tractable closed-form performance analysis in subsequent evaluations.

\setlength{\tabcolsep}{2.7pt}
\begin{table}[t!]
\centering
\caption{Table of Parameters \cite{zhang2024performance} ,\cite{shetty2023performance},\cite{fu2018performance}}
\label{table:parameters}
\scriptsize
\vspace{-3mm}
\begin{tabular}{|>{\centering\arraybackslash}p{1cm}|>{\centering\arraybackslash}p{1.2cm}|>{\centering\arraybackslash}p{1.35cm}|>{\centering\arraybackslash}p{1.35cm}|>{\centering\arraybackslash}p{1.12cm}|>{\centering\arraybackslash}p{1.5cm}|}
\hline
\multicolumn{1}{|c|}{{\textbf{Parameter}}} & \textbf{Value} & \textbf{Parameter} & \textbf{Value} & \textbf{Parameter} & \textbf{Value} \\
\hline
$\gamma_{th}$ & 15 \text{dB} & $w_0$ & 0.01 \text{m} & $\varepsilon_1$ & $0.03 \text{m}^{2} \cdot \text{s}^{-3}$ \\
\hline
$L_{sr}$ & 95m & $\lambda$ & 532 nm & $\varepsilon_2$ &  {{0.0005}$m^{2}.s^{-3}$} \\
\hline
$L_{rd}$ & 40m & $D_R$ & 0.05m & $A_S$ & $1.9 \times 10^{-4}$ \\
\hline
$\rho$ & 1 & $\omega$ & -2.5 & $A_T$ & $0.01863 $ \\
\hline
c & 0.15m$^{-1}$ & $\eta_{1,2}$ & $7.5 \times 10^{-5}$ & $A_{TS}$ & $9.41 \times 10^{-3}$ \\
\hline
$\sigma_\theta$ & $2 \times 10^{-3}$ & $\chi_{T_{1,2}}$ & $10^{-5} k^{2}.s^{-3}$ & $\sigma_{\tiny{S}}$(\tiny {Direct}) & 0.2 m \\
\hline
$\sigma_\phi$ &  $1.5 \times 10^{-3}$ & $P_{n}$\tiny{(active)} & -113 dBW  & $\sigma^2_n$(\tiny {O-RIS}) & -100 dBm \\
\hline
$L_D$ & 120 m &Gain\tiny{(active)}&   7 dB & $\sigma^2_n$(\tiny {Direct}) & -95.5 dBm \\
\hline
 $L_{\text{Fiber}}$  & $80 Km$ &$R_{\text{up}}$\tiny{(\text{POS-UOAP})}& 15 Gbps & $T_{ \text{propagation}}$  & \tiny{$5\times10^{\tiny-6}s/km$} \\
\hline
$\alpha_{Zipf} $ & $1$ & $T_{\text{processing}}$  & $0.2\times10^{-3} s$ & $T_{\text{FEC}}$  & $0.25\times10^{-3} s$ \\
\hline
\end{tabular} 
\end{table}

Fig.~2(b) compares the OP of a direct UWOC link and an O-RIS-assisted link across different \( \bar{\gamma} \) values under the OTOPS model~\cite{10897965} (see Appendix~B for more details). Despite the longer total distance in the O-RIS-assisted system with $L_{sr}=60\ \mathrm{m}$, $L_{rd}=40\ \mathrm{m}$, its performance surpasses that of the direct link with $L_{D}=90\ \mathrm{m}$, improving significantly with the addition of more O-RIS elements. For instance, at \( \bar{\gamma} = 40 \, \text{dB} \), the OP for a direct link is \( 2.43e-2 \), while for an O-RIS-assisted link, the OP is \( 8.36e-4 \) for \( N=4 \), and \( 2.73e-5 \) for \( N=16 \) under $T = 20^{\circ}\mathrm{C}, S = 30~\mathrm{ppt}, H = -0.2~\mathrm{^\circ C \, ppt^{-1}}$ and  $P = 0~\mathrm{dBar}$~\cite{10897965}. Furthermore, it can be observed that as temperature and salinity decrease, the $\sigma_I^2$ under strong oceanic turbulence reduces, thereby leading to a noticeable reduction in OP.

In this work, two models for the G--G model are presented in the appendices. Appendix A corresponds to moderate-to-strong turbulence conditions commonly adopted in previous studies, ensuring fair comparison with existing literature. In contrast, Appendix B introduces the OTOPS model, which accounts for temperature, salinity, and the oceanic turbulence spectrum, thereby offering a more physically grounded and realistic representation of underwater turbulence under weak-to-moderate conditions. The significance of the latter model lies in its enhanced accuracy under practical oceanic environments, making it a more reliable choice for real-world performance evaluation of UWOC systems.

Fig.~2(c) illustrates the OP under hybrid combining schemes. The results indicate that MRC consistently outperforms SC, with the performance gap widening at higher SNR levels. Furthermore, increasing the number of O-RIS elements substantially improves the effectiveness of both methods, leading to reduced OP and enhanced system reliability. 

\begin{figure*}[t!]
\centering
\renewcommand{\arraystretch}{0.95}

\setlength{\tabcolsep}{-5pt}
\begin{tabular}{@{}cccc@{}}
\includegraphics[width=0.249\textwidth]{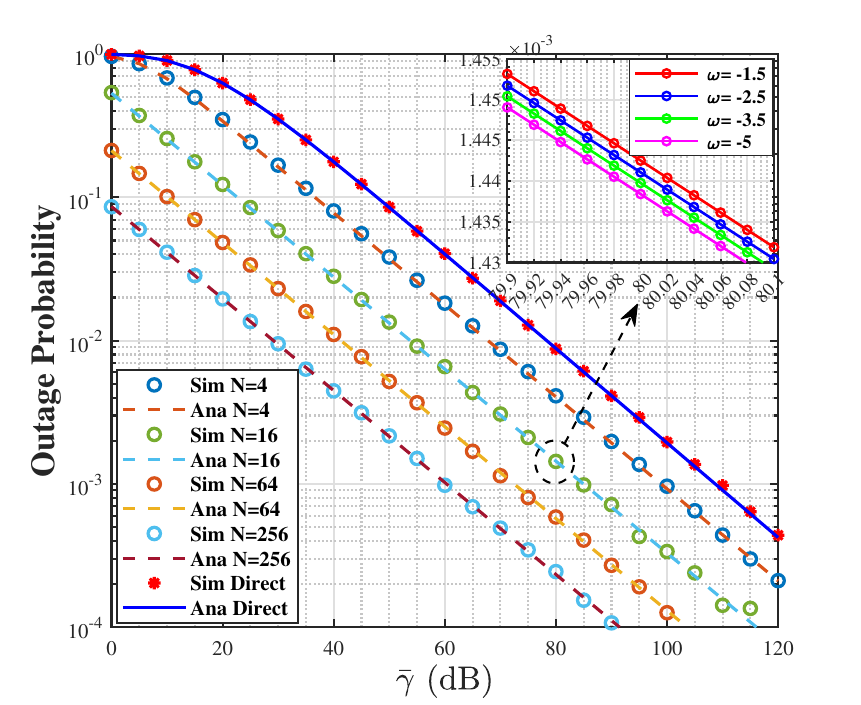} &
\includegraphics[width=0.249\textwidth]{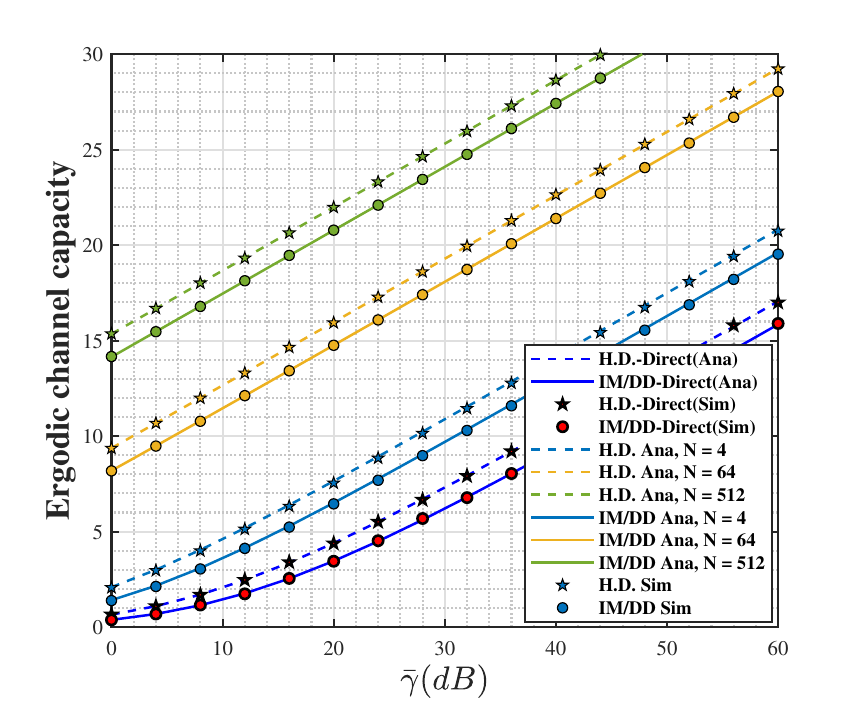} &
\includegraphics[width=0.249\textwidth]{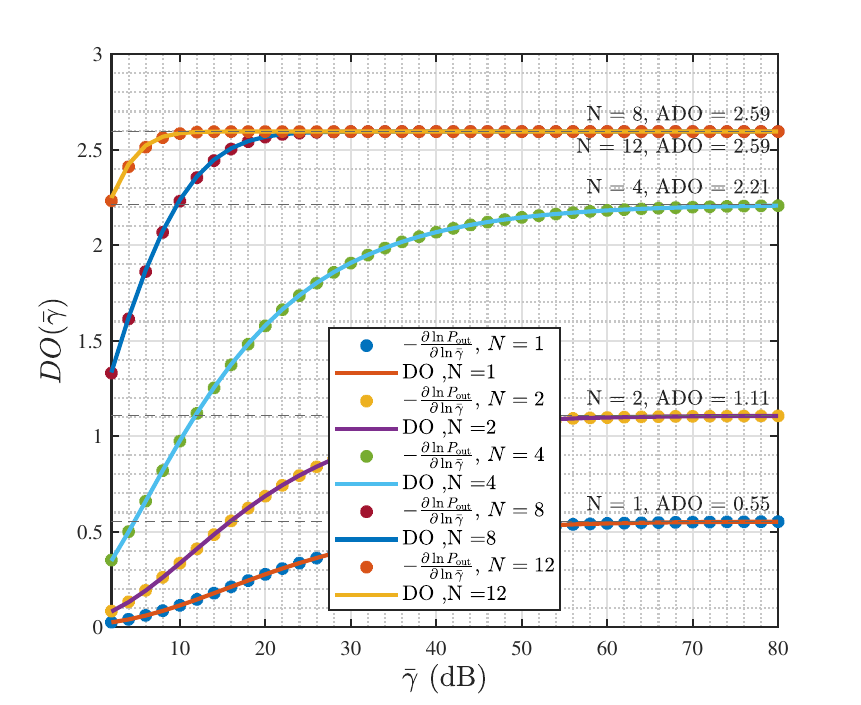} &
\includegraphics[width=0.249\textwidth]{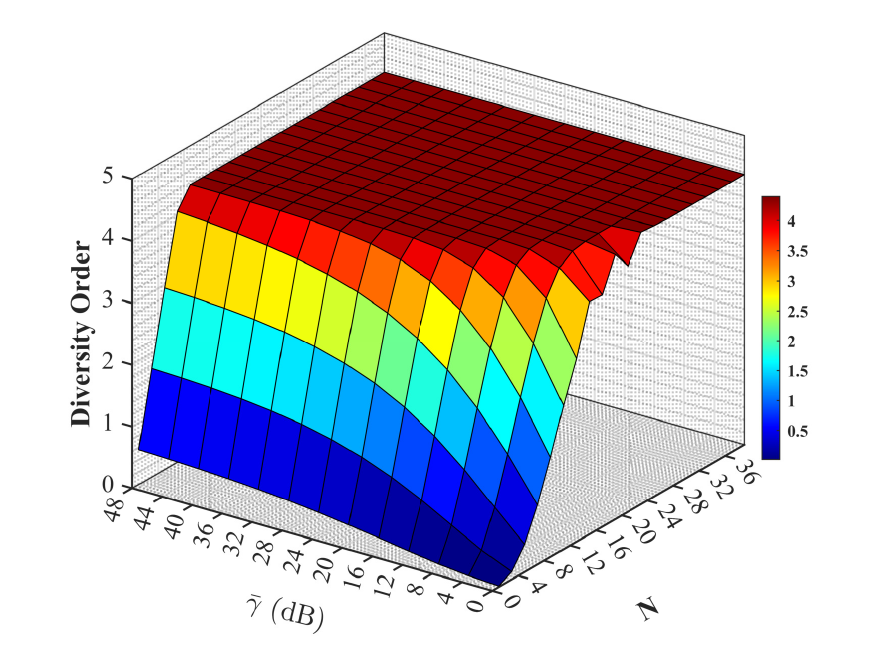} \\[-2pt]
\hspace{-2mm}\scriptsize\textbf{(a)} &
\scriptsize\textbf{(b)} &
\scriptsize\textbf{(c)} &
\scriptsize\textbf{(d)} \\
\end{tabular}

\vspace{0pt} 

\setlength{\tabcolsep}{-5pt}
\begin{tabular}{@{}cccc@{}}
\includegraphics[width=0.249\textwidth]{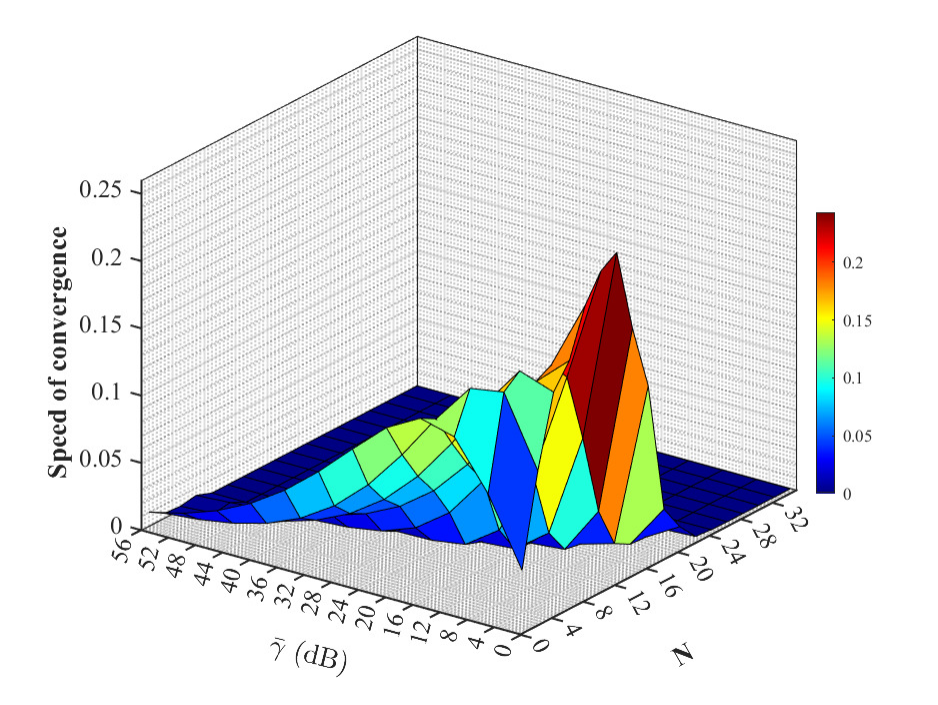} &
\includegraphics[width=0.249\textwidth]{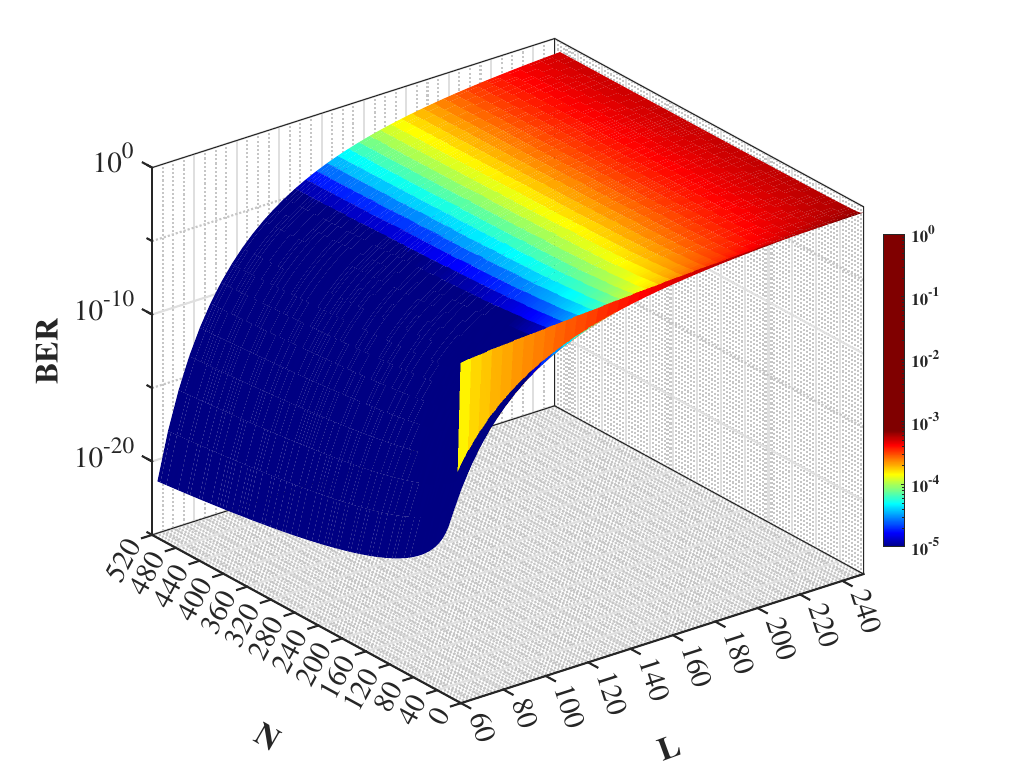} &
\includegraphics[width=0.249\textwidth]{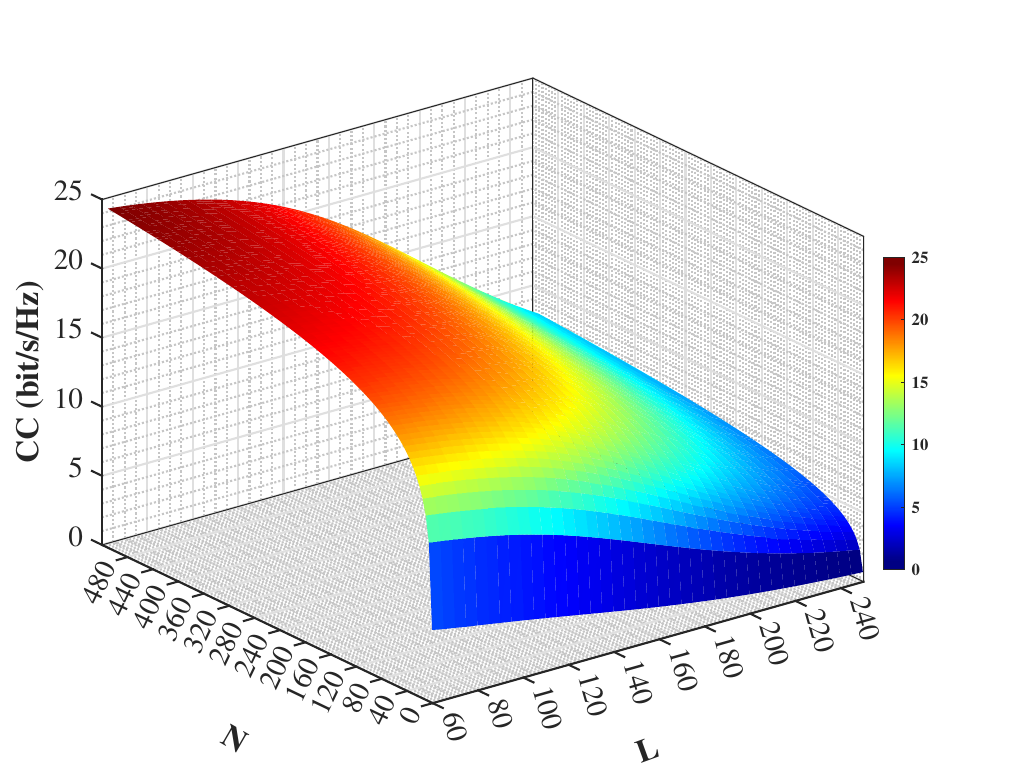} &
\includegraphics[width=0.249\textwidth]{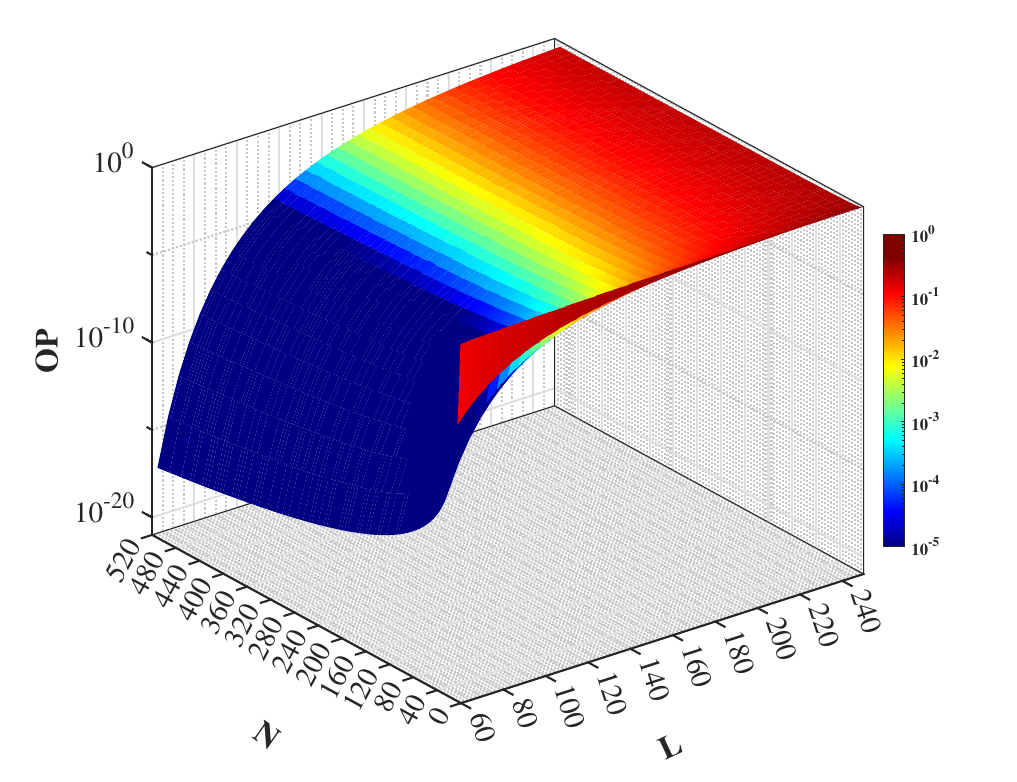} \\[-2pt]
\hspace{-2mm}\scriptsize\textbf{(e)} &
\scriptsize\textbf{(f)} &
\scriptsize\textbf{(g)} &
\scriptsize\textbf{(h)} \\
\end{tabular}

\caption{ 
(a) OP versus average SNR ($\bar{\gamma}$) for $L_{sr} = 90\,\mathrm{m}$, $L_{rd} = 60\,\mathrm{m}$, and $L_D = 140\,\mathrm{m}$. 
(b) Comparison of the CC under H.D. and IM/DD detection modes, considering a direct link with $L_D = 120$ m and an O-RIS–assisted link with $L_{sr} = 95$ m and $L_{rd} = 40$ m. 
(c) DO versus average SNR for $L_{sr} = L_{rd} = 60$ m. 
(d) 3D plot of DO as a function of $N_{\text{O-RIS}}$ and average SNR ($L_{sr} = L_{rd} = 40$ m). 
(e) 3D plot of SOC under the same setup. 
(f) 3D plot of BER versus $N_{\text{O-RIS}}$ and $L_{sd} \quad (L_{sr} = L_{rd})$. 
(g) 3D plot of CC versus $N_{\text{O-RIS}}$ and $L_{sd}$. 
(h) 3D plot of OP versus the same parameters.}

\label{fig:images_table2}
\end{figure*}

In Fig. 2(d), in contrast to the majority of prior works that consider only passive O-RIS configurations in UWOC systems, we conduct a more realistic performance evaluation of O-RIS using Monte Carlo (MC) simulation. The worst-case OP corresponds to a completely random passive setup, where $\rho_m \in (0,1]$ and $\theta_m \in [0, 2\pi)$. Subsequently, a passive configuration with phase control is considered, characterized by $\theta_m = -\angle(h_{\alpha,m}) + \epsilon_m$, where the phase error $\epsilon_m$ follows a Gaussian distribution with standard deviation $\sigma_{\text{ph.-error}} = \frac{\pi}{6}$. To reduce hardware complexity and constrain the phase error within a bounded range, we also evaluate a quantized phase control scheme, defined as $\theta_m = \text{Quantize}(-\angle(h_{\alpha,m}))$ with 3-bit resolution. The optimal passive performance, yielding the minimum OP, is achieved under the ideal conditions of $\rho_m = 1$ and $\theta_m = -\angle(h_{\alpha,m})$, which ensures perfect phase alignment. Finally, the lowest overall OP is observed in the active O-RIS case, where controlled gain is applied with a finer phase error tolerance, i.e., $\sigma_{\text{ph.-error}} = \frac{\pi}{36}$ (see Appendix G).

In Fig.~2(e), the effect of the modulation type is evident. Modulation schemes, including BPSK, BFSK, QPSK, and 16-QAM, have been evaluated and compared. Despite $L_{sd}$ exceeding $L_D$, the O-RIS--assisted link outperforms the direct link. Increasing the number of O-RIS elements reduces the error rate, whereas BPSK demonstrates superior performance compared to other modulation schemes.

Fig.~2(f) presents the analysis of the evaluated parameters with respect to the link length. The Meijer G-function in MATLAB was utilized, with the $\alpha$ and $\beta$ parameters precisely calculated to address the challenge of link-length-based derivation. The results were validated through MC simulation. Assuming $L_{sr}=L_{rd}$, results show that increasing link length degrades system performance. For lengths exceeding 150 meters, adding more O-RIS elements improves BER, as is the case for OP and CC, but the gain remains marginal.

In Fig.~2(g), the impact of varying the O-RIS location between the transmitter and receiver, simillar to Roller Linear Motion, was investigated across multiple scenarios with different link lengths (Tx--O-RIS and O-RIS--Rx) and water types (Pure Sea, Clear Ocean, Coastal Ocean with parameter $c(\lambda)$ = 0.056, 0.151, 0.305 (Tab. 3 of \cite{elfikky2024underwater})). Assuming a constant $\sigma_\phi$, changes in $L_{sr}$ and $L_{rd}$ affect $\sigma_\theta$ based on Eq.~(3) and Figs.~4 and 5 in~\cite{wang2021performance}. The results show that for shorter link lengths, system performance is optimal at the transmitter side. As the link length increases, system performance degrades due to significant path loss, causing signal arrival weak at the O-RIS. At the same BER, it is observed that as parameter c decreases, the optimal O-RIS position shifts closer to the receiver. This behavior is attributed to reduced absorption and scattering in clearer water. The influence of link length is further evident for link distances of 50 and 80 meters in the figure.

In Fig.~2(h), SOC is assessed under different jitter. As SNR increases, it converges toward zero, indicating saturation at the maximum achievable diversity order (ADO) under limited SNR. Higher jitter reduces the diversity order and accelerates convergence. The results for $N = 2, 4, 8$ align with the closed-form expression in~(52) of~\cite{zhang2024performance}, validating the analysis.

Fig.~3(a) compares the OP of a direct UWOC link and an O--RIS--assisted link across different \( \bar{\gamma} \) values, based on the model presented in Appendix~A for the G--G model. Similar to Fig.~2(b), despite the longer total distance in the O--RIS--assisted system, its performance surpasses that of the direct link and improves significantly with the addition of more O--RIS elements. For example, at \( \bar{\gamma} = 100 \, \text{dB} \), the OP for a direct link is \( 1.93e-3 \), while for an O-RIS--assisted link, the OP is \( 9.63e-4 \) for \( N=4 \), and \( 1.304e-4 \) for \( N=64 \). Furthermore, turbulence is modeled with the G–G distribution under the Appendix A model for the scintillation index. It shows that as $\omega$ decreases, meaning that temperature fluctuations increasingly dominate the oceanic turbulence, the scintillation index of light in strong oceanic turbulence decreases. As a result, the system's performance, including OP, is enhanced \cite{fu2018performance}.

Figure 3(b) examines the channel capacity for IM/DD and heterodyne receivers. With $L_{sr} = 95$ m, $L_{rd} = 40$ m, $L_D = 120$ m, $\sigma_\theta = 2 \times 10^{-3}$, and $\sigma_\phi = 1 \times 10^{-3}$, the O-RIS link achieves superior performance despite $L_{sd}$ being longer than $L_D$. Furthermore, as the number of O-RIS elements increases, the heterodyne receiver outperforms IM/DD, achieving more than one bit per second per hertz at a fixed SNR.

In Figure 3(c), we analyze the DO parameter. With \( L_{sr} = L_{rd} = 60 \, \text{m} \), \( \sigma_\theta = 1 \times 10^{-3} \), and \( \sigma_\phi = 0.5 \times 10^{-3} \), we observe that as the number of O-RIS elements increases, the Diversity Order quickly approaches the Asymptotic Diversity Order. The results align with the closed-form relation in eq. (42) of \cite{zhang2024performance}, confirming the accuracy of the obtained results.

In Figures 3(d) and 3(e), a 3D analysis of DO and SOC offers insights into O-RIS element selection and the optimal power range. The results show a peak in SOC for \( 15 < N < 24 \) and \( 4 < \bar{\gamma} < 20 \).
Finally, this study demonstrates that by utilizing advanced programming and optimization with the closed-form Meijer G-function in MATLAB, BER, CC, and OP were computed as functions of the number of O-RIS elements and link length in a large-scale system, considering $\bar{\gamma}$ = 25 dB, $\sigma_\theta = 2 \times 10^{-3}$, and $\sigma_\phi = 1.5 \times 10^{-3}$. Ultimately, Figures 3(f), (g), and (h) offer a comprehensive overview of the designed system, providing crucial insights into this field.

As illustrated in Fig.4(a), the end-to-end delay and per-UOAP throughput are presented for a UWOC- and O-RIS-enabled PON-based backhaul network, assuming negligible fiber noise. The horizontal axis represents the number of UOAPs, whereas the vertical axes indicate the delay (left) and throughput (right). The total latency experienced by each node consists of several components, including DBA latency, FEC operations, internal processing, queueing due to congestion, propagation delay, and retransmissions caused by packet loss. As the number of UOAPs increases, the delay curve gradually rises, mainly due to longer DBA cycles and intensified queueing effects. The throughput curves in ~Fig.4(a) show how the effective data rate is distributed among nodes under Zipf-like traffic demands~\cite{7493685}. The differences among minimum, maximum, and average throughput values result from load imbalance across nodes, whereas the ideal throughput curve highlights the reduction introduced by protocol overhead and congestion. An important observation emerges around the system saturation point. Considering the gross system capacity of $240~\mathrm{Gbps}$ and an efficiency factor of $85\%$, the effective usable capacity becomes $C_{\text{net}} = 240~\mathrm{Gbps} \times 0.85 = 204~\mathrm{Gbps}$. Dividing this net capacity by the physical ceiling of $15~\mathrm{Gbps}$ per UOAP yields $\mathfrak{T} = 204 / 15 \approx 13.6 \approx 14$ (rounded), which defines the critical saturation threshold.  

Pre-Saturation Region: 
For a limited number of UOAPs ($\mathfrak{T} \leq 4$), the effective network capacity exceeds the physical link limit. In this regime, the link ceiling acts as the primary throughput constraint, leading all UOAPs to saturate at nearly the same level. Consequently, the minimum, maximum, and average throughputs are almost identical, reflecting the pre-saturation behavior of the network under low user counts. Furthermore, for {\small{$\mathfrak{T} \leq 14$}}, the aggregate system capacity remains sufficiently high such that the fair-share bandwidth per user stays above the physical ceiling of $15~\mathrm{Gbps}$. In this region, both the ideal throughput (black dashed line with markers) and the realistic maximum throughput (green dashed line) converge to the upper bound of the physical port rate.

\hspace{-2mm}Post-Saturation Region and Throughput Analysis: For {\small{$\mathfrak{T} > 14$}}, the fair-share allocation falls below the $15~\mathrm{Gbps}$ ceiling as the number of users increases. Both the ideal throughput and the realistic maximum throughput decrease steadily with the number of UOAPs, following similar downward trends. However, the ideal throughput consistently remains slightly higher than the realistic maximum throughput. This difference arises from protocol and system overheads: the ideal throughput assumes perfect efficiency, neglecting overheads, error-control mechanisms, and other practical inefficiencies. In contrast, the realistic maximum throughput accounts for these overheads while simultaneously benefiting from traffic asymmetry induced by the Zipf distribution~\cite{7493685} and dynamic bandwidth allocation. As a result, high-demand users may achieve more than their nominal fair share by utilizing residual capacity left unused by low-demand users. The difference between ideal and realistic throughput can also be understood in terms of gross versus net system capacity. The gross capacity corresponds to the maximum physical data rate of 
$15~\mathrm{Gbps}$ per UOAP (i.e., $240~\mathrm{Gbps}$ system-wide), 
whereas the net capacity accounts for protocol overheads and other inefficiencies. 
Considering an efficiency factor of $85\%$, the effective usable throughput is 
obtained as $240~\mathrm{Gbps} \times 0.85 = 204~\mathrm{Gbps}$. This distinction explains the smoother, nonlinear decline observed in the average throughput curve under realistic conditions and highlights the combined influence of physical limitations, protocol overhead, and traffic imbalance on network performance.  

Average Throughput Behavior: The average throughput curve (red solid line) represents the effective capacity across all ONUs under heterogeneous traffic. Because the Zipf distribution concentrates heavy demand on a small subset of users, the average throughput consistently remains below the ideal fair-share case. This highlights the combined effects of traffic imbalance, protocol overhead, and physical constraints.  

\begin{figure}[!t]
\raggedright
\begin{minipage}{0.48\textwidth}
\centering
\hspace{-1.5mm}
\includegraphics[width=0.5\columnwidth, height=3.85cm, trim={12pt 0 2pt 22pt}, clip]{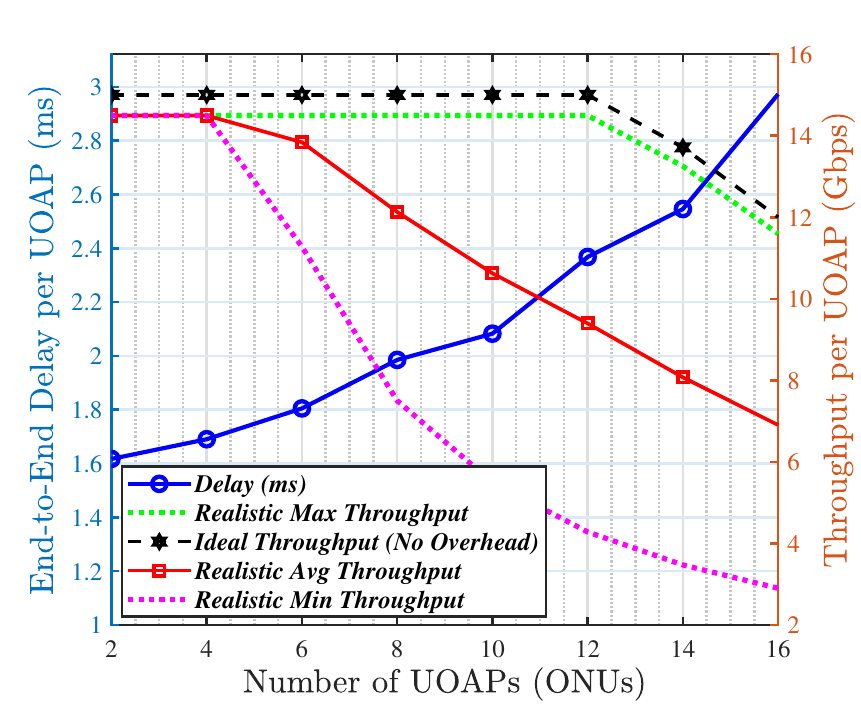}%
\hspace{0mm}%
\includegraphics[width=0.495\columnwidth, height=3.9cm, trim={15pt 0 30pt 20pt}, clip]{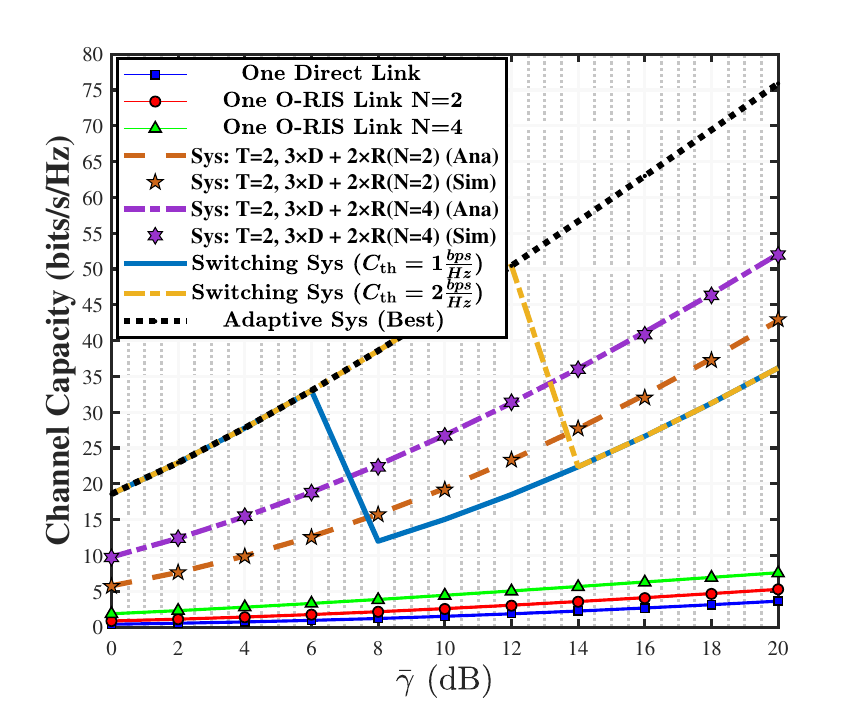}\\[-4pt]
\scriptsize
\makebox[0.47\columnwidth][c]{\textbf{(a)}}%
\hspace{1mm}
\makebox[0.47\columnwidth][c]{\hspace{3mm}\textbf{(b)}}
\vspace{-3mm}
\caption{(a) E2E delay and throughput allocated to each UOAP vs. the number of UOAPs. (b) CC of the IM/DD for $\mathfrak{T}=2$ under hard switching, where the direct links $L_D$ are $90\,\mathrm{m}$ and the O-RIS links $L_{SR}=L_{RD}$ are $50\,\mathrm{m}$.}
\end{minipage}
\end{figure}

These findings demonstrate that modern PONs, leveraging adaptive bandwidth allocation, can effectively accommodate heterogeneous user demands and achieve higher efficiency compared to rigid fair-share allocation. This adaptability represents a significant advantage of next-generation backhaul networks operating under diverse traffic loads (see Appendix~E).  

In Fig. 4(b), the overall system capacity under IM/DD is evaluated for two UOAPs connected to a fiber-optic backhaul, each employing three direct links and two O-RIS--assisted links. To enhance system performance, a hard-switching technique is applied to all links within each UOAP. In this approach, a direct link is selected if its capacity exceeds a predefined threshold; otherwise, an O-RIS--assisted link is used. In the optimal case, for each SNR value, the link offering the highest capacity among all available options is chosen (see Subsection~C of Appendix~E for more details).

\section{Conclusion and Future Work} 
This study presents a novel architecture combining fiber optic backhaul, UOAP, and O-RIS to achieve wide coverage, high speed, and low latency in UWOC by creating NLOS paths that outperform direct links. Passive O-RIS (ideal/quantized) provides significant gains with minimal energy, whereas active O-RIS achieves superior performance at higher energy and complexity costs. MRC consistently outperforms SC, particularly at high SNRs and with more O-RIS elements. Placing a passive O-RIS near the receiver reduces performance beyond the midpoint in the linear regime due to signal attenuation. Scalability is achieved via distributed UOAPs and fiber backhaul, ensuring high throughput with minimal delay impact. The models are validated through MC simulation; however, future work requires real-world testing, UOAP placement optimization, environmental adaptation, and energy minimization.

\vspace{2mm}
To complement the main discussions, a set of appendices is included at the end of this paper, containing detailed technical derivations and supplementary explanations.

\section{Appendix A}
Within the G–G turbulence model, and under the assumption of plane wave propagation, the turbulence parameters $\alpha$ and $\beta$ are calculated as follows:

{\small
\begin{align}
\alpha &= \left[ \exp \left( \frac{0.17 \, \sigma_{l_1}^2}{\left(1 + 0.167 \, \sigma_{l_1}^{12/5}\right)^{7/6}} \right) - 1 \right]^{-1},\label{eq12} \\
\beta &= \left[ \exp \left( \frac{0.225 \, \sigma_{l_2}^2}{\left(1 + 0.259 \, \sigma_{l_2}^{12/5}\right)^{5/6}} \right) - 1 \right]^{-1}. \label{eq13}
\end{align} }

According to \cite{fu2018performance, R29, R30, zhang2024performance, naik2022evaluation}, $\sigma_{l_i}^2$ (for $i=1,2$) denotes the Rytov variance of the two paths, which can be expressed by Equation (3) of \cite{fu2018performance}. In strong underwater turbulence, the Rytov variance satisfies $0 \le \sigma_{l_i}^2 \le \infty$, and its relationship is given in the following form:
\begin{equation}
\sigma_l^2 = 1.23 C_n^2 k^{7/6} L^{11/6},
\end{equation}

where $k = \frac{2\pi}{\lambda}$, $\lambda$ is the optical wavelength, $L$ is the link distance, and $C_n^2$ is the refractive index structure constant.

Under moderate-to-strong turbulence conditions, it is defined as:
\begin{align}
\centering 
C_n^2 &= 16 \pi^2 k^{-7/6} L^{-11/6} \Re \Bigg\{ 
\int_{0}^{L} d\zeta \int_0^{\infty} \kappa \Big[ 
E(\zeta, \kappa, L) \\
&\quad \times E(\zeta, -\kappa, L) \notag  + |E(\zeta, \kappa, L)|^2 \Big] \Phi_n(\kappa) \, d\kappa 
\Bigg\}.
\end{align}

In this expression, $\Re\{\cdot\}$ denotes the real part, and $E(\zeta, \kappa, L) = i k \exp \Big[-0.5 i \zeta (L - \zeta) \kappa^2 / (k L) \Big],$
where $\kappa$ is the magnitude of the spatial frequency, and $\Phi_n(\kappa)$ is the power spectrum of homogeneous and isotropic oceanic water. Under the assumption of equal eddy thermal diffusivity and the diffusion of salt, it is given by \cite{fu2018performance}:
\begin{align}
\Phi_n(\kappa) &= 0.388 \times 10^{-8} \epsilon^{-1/3} \kappa^{-11/3}(1+ 2.35 \kappa^{2/3} u^{1/2} \epsilon^{-1/6}) \notag \\
&\quad \omega^{-2} X_T (\omega^2 e^{-A_T \delta} + e^{-A_S \delta} - 2 \omega e^{-A_{TS} \delta})
\end{align}
where $X_T$ is the rate of dissipation of the mean square temperature, $\epsilon$ is the rate of dissipation of the kinetic energy per unit mass of fluid, $u$ is the kinematic viscosity and $\omega$ is a unitless parameter providing the ratio of temperature contributions to salinity to the refractive index spectrum. When ocean turbulence is temperature and salinity induced, $\omega$ ranges from $-5$ to $0$. Moreover, $A_S = 1.9 \times 10^{-4}$, $A_T = 1.863 \times 10^{-2}$, $A_{TS} = 9.41 \times 10^{-3}$, and $\delta = 8.28 \kappa^{4/3} u \epsilon^{-1/3} + 12.978 \kappa^2 u^{3/2} \epsilon^{-1/2}$.

First, Eq. (15) is numerically evaluated, then $\sigma_l^2$ is obtained from Eq. (14), and by inserting $\sigma_l^2$ into \eqref{eq12} and \eqref{eq13}, the parameters $(\alpha, \beta)$, which represent the \textit{small-scale} and \textit{large-scale} scintillation indices, respectively, can be calculated. One of the challenges in simulating the results presented in the study was the accurate calculation of these parameters for various link lengths, as well as the successful implementation and simulation of these parameters in MATLAB.
In the proposed turbulence model, disturbances caused by salinity, temperature, and other environmental factors in the marine environment are represented through mathematical equations. Furthermore, parameters such as $\epsilon$ (the rate of kinetic energy dissipation), $\nu$ (kinematic viscosity), and $X_r$ (the rate of mean-square temperature dissipation), along with $\omega$, are incorporated into the model, as they relate to the physical properties of the medium and fluid dynamics. By integrating these parameters, the model effectively captures turbulence originating from various sources (temperature, salinity, and other physical factors) within a unified mathematical framework, which is used to calculate the light intensity variance $\sigma_i^2$ in the marine environment. The advantage of this model lies in its comprehensiveness, as it simultaneously accounts for multiple factors influencing light propagation in aquatic environments.

\section{Appendix B} 
\begin{center}
Theoretical Basis and Computation of Optical Parameters in the OTOPS Model
\end{center}

The significance of this subject has been thoroughly examined in~\cite{10897965},~\cite{ata2023analysis}, and~\cite{ata2021ber}, where various aspects of turbulence-induced optical fluctuations have been analyzed.

\subsection{Rytov Variance of Gaussian Beam in Weak Turbulence}

Under weak-to-moderate turbulence conditions, the Rytov variance, which quantifies the log-amplitude fluctuations of a Gaussian beam and is equivalent to the scintillation index, can be expressed as~\cite{10897965,Andrews2005}:
\begin{equation}
\begin{aligned}
\sigma_B^2 &= 8 \pi^2 k^2 L 
\int_0^1 \mathrm{d}\xi 
\int_0^\infty \kappa \, \Phi_n(\kappa) \, 
\exp\!\Big(-\frac{\Lambda_1 L k^2 \xi^2}{k}\Big) \\
&\quad \times 
\Bigg\{ 1 - \cos\Bigg[ \frac{L k^2}{k} \, \xi \, (1 - \overline{\Theta}_1 \xi) \Bigg] \Bigg\} 
\, \mathrm{d}\kappa,
\end{aligned}
\label{eq:A1}
\end{equation}
where $\xi$ denotes the normalized propagation path, $\kappa$ is the spatial wavenumber magnitude, $k = 2\pi/\lambda$ represents the optical wavenumber, and $\Lambda_1 = \Lambda_0 / (\Theta_0^2 + \Lambda_0^2)$ corresponds to the Fresnel ratio of the Gaussian beam at the receiver.

\subsection{OTOPS Turbulence Power Spectrum}

The Oceanic Turbulence Optical Power Spectrum (OTOPS) model~\cite{Yao2020} provides a realistic description of underwater refractive-index fluctuations. As described by this model, the power spectral density of refractive index variations is formulated as~\cite{10897965}:
\begin{equation}
\Phi_n(\kappa) = A^2 \Phi_T + B^2 \Phi_S + 2AB \Phi_{TS},
\label{eq:A2}
\end{equation}
where $A = \partial n(T,S,\lambda)/\partial T$ and $B = \partial n(T,S,\lambda)/\partial S$ represent the sensitivity coefficients of the refractive index $n(T,S,\lambda)$ with respect to temperature and salinity, respectively.
Following the empirical model introduced by Quan and Fry~\cite{Quan1995}, the refractive index of seawater is given by~\cite{ata2021ber}:
\begin{equation}
\begin{aligned}
n(T,S,\lambda) =\,& a_0 + (a_1 + a_2 T + a_3 T^2)S + a_4 T^2 \\
& + (a_5 + a_6 S + a_7 T)\lambda^{-1} + a_8 \lambda^{-2} + a_9 \lambda^{-3},
\end{aligned}
\label{eq:4}
\end{equation}
where the coefficients are $a_0 = 1.31405$, $a_1 = 1.779 \times 10^{-4}$, $a_2 = -1.05 \times 10^{-6}$, $a_3 = 1.6 \times 10^{-8}$, $a_4 = -2.02 \times 10^{-6}$, $a_5 = 15.868$, $a_6 = 0.01155$, $a_7 = -0.00423$, $a_8 = -4382$, and $a_9 = 1.1455 \times 10^{6}$. Here, $T$ and $S$ denote the temperature and salinity, respectively.

As described in~\cite{Yao2020}, the mean refractive index for average temperature and salinity values is obtained as
\begin{equation}
\begin{aligned}
&n_0(\langle T \rangle, \langle S \rangle, \lambda) =\, 
a_0 + (a_1 + a_2 \langle T \rangle + a_3 \langle T \rangle^2)\langle S \rangle 
+ a_4 \langle T \rangle^2 \\
& + (a_5 + a_6 \langle S \rangle + a_7 \langle T \rangle)\lambda^{-1} 
+ a_8 \lambda^{-2} + a_9 \lambda^{-3}.
\end{aligned}
\label{eq:5}
\end{equation}

The temperature and salinity coefficients $A$ and $B$ are derived from the partial derivatives of $n(T,S,\lambda)$ as follows:
\begin{equation}
\begin{aligned}
A(\langle T \rangle, \langle S \rangle, \lambda) 
=\, & \left. \frac{\partial n(T,S,\lambda)}{\partial T} \right|_{T=\langle T \rangle,\, S=\langle S \rangle} \\
=\, & a_2 \langle S \rangle + 2 a_3 \langle T \rangle \langle S \rangle 
+ 2 a_4 \langle T \rangle + a_7 \lambda^{-1},
\end{aligned}
\label{eq:6}
\end{equation}
\begin{equation}
\begin{aligned}
B(\langle T \rangle, \langle S \rangle, \lambda) 
=\, & \left. \frac{\partial n(T,S,\lambda)}{\partial S} \right|_{T=\langle T \rangle,\, S=\langle S \rangle} \\
=\, & a_1 + a_2 \langle T \rangle + a_3 \langle T \rangle^2 + a_6 \lambda^{-1}.
\end{aligned}
\label{eq:7}
\end{equation}

The power spectra corresponding to temperature, salinity, and their cross-spectral components are expressed as:
\begin{equation}
\begin{aligned}
\Phi_i(\kappa) &= 
\frac{\beta_0 \varepsilon^{-1/3} \kappa^{-11/3} \chi_i }
     {4\pi}  \times \exp\!\Big[-174.9 (\kappa \eta)^2 c_i^{0.96}\Big]\\
& \Big[ 1 + 21.61 (\kappa \eta)^{0.61} c_i^{0.02} -18.18 (\kappa \eta)^{0.55} c_i^{0.04}\Big],
\end{aligned}
\label{eq:A3}
\end{equation}
where $i \in \{T, S, TS\}$, $\beta_0 = 0.72$, $\varepsilon$ is the turbulent kinetic energy dissipation rate, $\chi_T$ is the temperature variance dissipation rate, and $\chi_S = d_r \chi_T / H^2$, $\chi_{TS} = 0.5(1 + d_r)\chi_T / H$ represent the dissipation rates for salinity and co-spectral components, respectively.

\subsection{Eddy Diffusivity Ratio and Gradient Effects}

The coupling between temperature and salinity fluctuations is governed by the eddy diffusivity ratio $d_r$, defined as~\cite{Elamassie2017}:
\begin{equation}
d_r =
\begin{cases}
R_p + \sqrt{R_p}\sqrt{R_p - 1}, & R_p \ge 1,\\[4pt]
1.85 R_p - 0.85, & 0.5 \le R_p < 1,\\[4pt]
0.15 R_p, & R_p < 0.5,
\end{cases}
\label{eq:A5}
\end{equation}
where the density ratio $R_p$ is determined by
\begin{equation}
R_p = \frac{|H| \, \alpha_T}{\beta_S},
\label{eq:Rp}
\end{equation}
with $H = \partial T / \partial S$ representing the  temperature-salinity gradient ratio. The thermal expansion and haline contraction coefficients, $\alpha_T$ and $\beta_S$, are the
thermal expansion coefficient and the saline contraction coefficient~\cite{Yao2020}, are computed using the \textit{Gibbs SeaWater (GSW)} toolbox (TEOS-10 toolbox) as~\cite{McDougallBarker2020}:
\begin{equation}
\alpha_T = gsw\_alpha(S_A, C_T, p), \quad
\beta_S = gsw\_beta(S_A, C_T, p).
\end{equation}
\subsection{Non-dimensional Coefficients}
The dimensionless empirical coefficients $c_i$ used in the spectral model are defined as~\cite{ata2021ber}:
\begin{equation}
\begin{aligned}
c_T &= 0.072^{4/3}\beta_0 P_r^{-1}, \\
c_S &= 0.072^{4/3}\beta_0 S_c^{-1}, \\
c_{TS} &= 0.072^{4/3}\beta_0 \frac{P_r + S_c}{2 P_r S_c},
\end{aligned}
\label{eq:A4}
\end{equation}
where $P_r$ and $S_c$ denote the Prandtl and Schmidt numbers, respectively.

\subsection{Statistical Characterization of Log-Amplitude Fluctuations}
The parameters $\sigma^2_{\ln X}$ and $\sigma^2_{\ln Y}$ represent the influence of large- and small-scale turbulent eddies on the propagating optical wave and can be expressed as~\cite{Andrews2005,10897965};
\begin{equation}
\sigma^2_{\ln X} = 
\dfrac{0.49\,\sigma_B^2}
{\left[1 + 0.56(1+\Theta_1)\sigma_B^{12/5}\right]^{7/6}}, \label{eq:A5}
\end{equation}
\begin{equation}
\sigma^2_{\ln Y} =
\dfrac{0.51\,\sigma_B^2}
{\left(1 + 0.69\sigma_B^{12/5}\right)^{5/6}},\label{eq:A6}
\end{equation}
where $\sigma_B^2$ denotes the Rytov variance of a Gaussian beam. A detailed analytical formulation of $\sigma_B^2$ is provided in ~\eqref{eq:A1}. The quantity $\Theta_1 = \Theta_0 / (\Theta_0^2 + \Lambda_0^2)$ [m] characterizes the curvature parameter of the beam at the receiver, with $\Theta_0$ and $\Lambda_0$ describing the beam curvature parameters at the transmitter, respectively.

The term $\Lambda_0 = 2L/(kW_0^2)$ depends on the beam radius $W_0$ [m] and the propagation distance $L$, while $\Theta_0 = 1 - L/F_0$ defines the curvature parameter at the transmitter based on the phase-front radius $F_0$ [m]. In other words, $F_0$ determines the radius of the phase front curvature of the optical wave. 

To accurately characterize optical propagation in realistic underwater environments, the Oceanic Turbulence Optical Power Spectrum model is adopted, as it provides a physically consistent representation of turbulence effects by incorporating practical water parameters \cite{10897965}. 

\subsection{Relation to Model Implementation}

In the proposed implementation, the parameters of the \textit{Gamma--Gamma} distribution, denoted as $\alpha_{GG}$ and $\beta_{GG}$, are determined from the logarithmic amplitude variances $\sigma_{\ln X}^2$ and $\sigma_{\ln Y}^2$, respectively:
\begin{equation}
\alpha_{GG} = \frac{1}{e^{\sigma_{\ln X}^2} - 1}, \qquad
\beta_{GG} = \frac{1}{e^{\sigma_{\ln Y}^2} - 1}.
\end{equation}
These variances are expressed as functions of the Rytov variance $\sigma_B^2$, which itself is obtained from~\eqref{eq:A1} using the OTOPS spectral formulation given in~\eqref{eq:A2}--\eqref{eq:A6}. To preserve accuracy and support model verification, several auxiliary quantities are also computed, including $\chi_S$, $\chi_{TS}$, $\sigma_B^2$, $\sigma_{\ln X}^2$, $\sigma_{\ln Y}^2$, and $\sigma_I^2$.

\subsection{Summary of Model Parameters}

Table~\ref{tab:otops_params} summarizes the physical, optical, and turbulence-related 
parameters utilized in the OTOPS model. Each parameter is listed along with its 
definition and corresponding units for clarity and reference.

\begin{table}[h!]
\centering
\caption{Summary of Parameters Used in the OTOPS Model}
\renewcommand{\arraystretch}{1.15}
\begin{tabular}{lll}
\hline
\textbf{Symbol} & \textbf{Description} & \textbf{Units} \\
\hline
$S_A$ & Absolute salinity & g/kg \\
$C_T$ & Conservative temperature & $^\circ$C \\
$p$ & Pressure & dbar \\
$\lambda$ & Optical wavelength & m \\
$L_{sr}, L_{rd}$ & Propagation path lengths & m \\
$\varepsilon$ & TKE dissipation rate & m$^2$/s$^3$ \\
$\chi_{_{T}}$ & Temperature variance dissipation rate & K$^2$/s \\
$H$ & Temperature–salinity gradient ratio & K/m \\
$\beta_0$ & Empirical OTOPS constant & --- \\
$D_G$ & Beam diameter & m \\
$W_0$ & Beam waist & m \\
$F_0$ & Focal length & m \\
$k$ & Wave number: $2\pi/\lambda$ & rad/m \\
$\alpha_T$ & Thermal expansion coefficient & 1/$^\circ$C \\
$\beta_S$ & Haline contraction coefficient & 1/(g/kg) \\
$R_p$ & Density ratio $|H|\alpha_T/\beta_S$ & --- \\
$d_r$ & Eddy diffusivity ratio & --- \\
$\alpha_{GG}, \beta_{GG}$ & Gamma-Gamma parameters & --- \\
\hline \label{tab:otops_params}
\end{tabular}
\end{table}

\section{Appendix C}
The expressions for the outage probability using the mEGG model in this study follow the same analytical framework as in \cite{10413214}, with the necessary adaptations for our analysis, as outlined in the relevant section. The probability density function (PDF) and cumulative distribution function (CDF) for the instantaneous SNR of the UWOC link under mEGG turbulence, considering various underwater turbulence conditions, are analyzed using the mEGG distribution for the source-to-destination link. Hence, the PDF of $\gamma$ is written as:
\begin{equation}
f_{\gamma_{mEGG}}(\gamma) = \Xi_1 \gamma^{\frac{m_u}{r}-1} 
G^{1,0}_{0,1}\!\left[\Xi_2 \gamma^{1/r}\,\middle|\,
\begin{matrix}
- \\ 0
\end{matrix}
\right]
\end{equation}
where {\small
\begin{equation*}
\begin{aligned}
 &\Xi_1=\frac{\big(E(Y_u)\big)^{m_u}}{r\,\Gamma(m_u)\,w_u^{m_u}\,\mu_r^{m_u/r}}, \Xi_2 = \frac{E(Y_u)}{w_u \mu_r^{1/r}},
m_u = \frac{\big(E(Z_u)\big)^2}{\mathrm{Var}(Z_u)},\\
& w_u = \frac{\mathrm{Var}(Z_u)}{E(Z_u)},
E(Z_u) = N_U E(Y_u),
\mathrm{Var}(Z_u) = N_U \,\mathrm{Var}(Y_u).
\end{aligned}
\end{equation*} }
Here, $r$ represents the detection techniques used (that is, $r=1$ corresponds to Heterodyne Detection (HD) and $r=2$ corresponds to intensity modulation / direct detection (IM / DD)), and $\mu_r$ is the electrical SNR. 
 
 Furthermore, the CDF of $\gamma$ is obtained as:
\begin{equation}
F_{\gamma_{mEGG}}(\gamma) = \Xi_3 \gamma^{m_u/r} 
G^{r,1}_{1,r+1}\!\left[\Xi_4 \gamma \,\middle|\,
\begin{matrix}
1-m_u/r \\
0,\Xi_5,-m_u/r
\end{matrix}
\right],
\end{equation}
and the outage probability is given by
\begin{equation}
OP_{mEGG}(\gamma) = F_{\gamma_{mEGG}}(\gamma_{_{th}}).
\end{equation}
where { 
$
\Xi_3 = \frac{(E(Y_u))^{m_u}}{\sqrt{r}(2\pi)^{\tfrac{r-1}{2}}\Gamma(m_u)w_u^{m_u}\mu_r^{m_u/r}},\;
\Xi_4 = \frac{(E(Y_u))^r}{w_u^r \mu_r r^r},\;
\Xi_5 = \tfrac{r-1}{r},\;
\Xi_6 = \omega_s \omega_r \xi_{u,s}^2 \xi_{u,r}^2,\;
\Xi_7 = A_{u,s} A_{u,r} \lambda_s \lambda_r,\;
\Xi_8 = \frac{(1-\omega_r)\omega_s \xi_{u,s}^2 \xi_{u,r}^2}{c_r^{1/2} (2\pi)^{\frac{1}{2}(c_r-1)}\Gamma(a_r)},\;
\Xi_9 = (\lambda_s b_r c_r A_{u,s} A_{u,r})^{c_r},\;
\Xi_{10} = 0,\ldots,\tfrac{c_r-1}{c_r},\;
\Xi_{11} = \tfrac{1-\xi_{u,s}^2}{c_r},\ldots,\tfrac{c_r-\xi_{u,s}^2}{c_r},\;
\Xi_{12} = -\tfrac{\xi_{u,s}^2}{c_r},\ldots,\tfrac{c_r-\xi_{u,s}^2 -1}{c_r},\;
\Xi_{13} = \frac{(1-\omega_s)\omega_r \xi_{u,s}^2 \xi_{u,r}^2}{c_s^{1/2} (2\pi)^{\frac{1}{2}(c_s-1)}\Gamma(a_s)},\;
\Xi_{14} = \frac{1}{(\lambda_r b_s c_s A_{u,r} A_{u,s})^{c_s}},\;
\Xi_{15} = \tfrac{\xi_{u,r}^2 +1}{c_s},\ldots,\tfrac{c_s+\xi_{u,r}^2}{c_s},\;
\Xi_{16} = \tfrac{1}{c_s},\ldots,1,\;
\Xi_{17} = \tfrac{\xi_{u,r}^2}{c_s},\ldots,\tfrac{c_s-1+\xi_{u,r}^2}{c_s},\;
\Xi_{18} = \frac{(1-\omega_s)(1-\omega_r)\xi_{u,s}^2 \xi_{u,r}^2}{\Gamma(a_s)\Gamma(a_r)},
\Xi_{19} = (b_s A_{u,s})^{c_s}(b_r A_{u,r})^{c_r}
$. }

\vspace{2mm}
The $k$-th moment of $Y_u$ is given by \cite[Eq.(7.811.4)]{gradshteyn2014table}:
\begin{equation}
\begin{aligned}
&E(Y_u^k) = \frac{\Xi_6 \Xi_7^k \prod_{j=1}^{4}\Gamma(B_{1j}+k)}{\prod_{j=1}^{2}\Gamma(D_{1j}+k)} 
+ \frac{\Xi_8 \Xi_9^k \prod_{j=1}^{4}\Gamma(B_{2j}+k)}{\prod_{j=1}^{2}\Gamma(D_{2j}+k)} \\[6pt]
&\quad + \frac{\Xi_{13} \Xi_{14}^k \prod_{j=1}^{4}\Gamma(B_{3j}+k)}{\prod_{j=1}^{2}\Gamma(D_{3j}+k)} 
+ \frac{\Xi_{18} \Xi_{19}^k \prod_{j=1}^{4}\Gamma(B_{4j}+k)}{\prod_{j=1}^{2}\Gamma(D_{4j}+k)}.
\end{aligned}
\end{equation}

where 
\vspace{2mm}

\noindent $ B_{1j} = [1,\, \xi_{u,r}^2,\, 1,\, \xi_{u,s}^2],\;
B_{2j} = \big[a_r,\, \tfrac{\xi_{u,r}^2}{c_r},\, 1-\Xi_{10},\, 1-\Xi_{11}\big],\;
B_{3j} = \big[a_s,\, \tfrac{\xi_{u,s}^2}{c_s},\, \Xi_{16},\, \Xi_{17}\big],\;
B_{4j} = \big[a_r\tfrac{\xi_{u,r}^2}{c_r},\, a_s\tfrac{\xi_{u,s}^2}{c_s}\big],\;
D_{1j} = [\xi_{u,r}^2+1,\, \xi_{u,s}^2+1],\;
D_{2j} = [\tfrac{\xi_{u,r}^2}{c_r}+1,\, 1-\Xi_{12}],\;
D_{3j} = [\tfrac{\xi_{u,s}^2}{c_s}+1,\, \Xi_{15}],\;
D_{4j} = [\tfrac{\xi_{u,r}^2}{c_r}+1,\, \tfrac{\xi_{u,s}^2}{c_s}+1].
$

\vspace{2mm}
In the mEGG turbulence-based outage probability model, it is assumed that the pointing error parameter $\xi$ and $A_u$ are identical to those used in the G–G model (for more details, see Lemma 2 in Subsection C of Section II in \cite{10413214}).
The mEGG parameters are taken from Table 1 in \cite{8606206}, which is considered a reliable reference for experimentally validated parameters in this context, and are further provided in Table~\ref{tab:mEGG}.

\section{Appendix D}
Considering (1) for the \textit{passive} case, the overall fading, expressed as the product of turbulence fading $(h_\alpha)$ and pointing error fading $(h_P)$, that is, $h = h_\alpha \cdot h_P$, can be characterized by the PDF of each component for both the direct links and the O-RIS–assisted links, which is provided here for clarity based on the studies in \cite{naik2022evaluation, zhang2024performance, R26, R27, 10793105, R28}:

\begin{table}[t!]
\centering
\caption{\scriptsize Measured and estimated parameters of the EGG distribution along with the goodness of fit tests \\ for the gradient temperature UWOC system~\cite{8606206}.}
\label{tab:mEGG}
\renewcommand{\arraystretch}{1.65} 
\resizebox{\columnwidth}{!}{%
\begin{tabular}{|c|c|c|c|c|c|c|}
\hline
{\begin{tabular}[c]{@{}c@{}}\textbf{Bubbles Level}\\ \textbf{BL (L/min)}\end{tabular}} &
{\begin{tabular}[c]{@{}c@{}}Temperature\\ Gradient\\ ($^\circ$C.cm$^{-1}$)\end{tabular}} &
{$\sigma^2_{I,\text{meas}}$} &
\multicolumn{4}{c|}{\textbf{Exponential–Generalized Gamma Distribution}} \\ \cline{4-7}
 & & & $\sigma^2_I$ & $(\omega,\lambda,a,b,c)$ & MSE & $R^2$ \\ \hline
2.4 & 0.05 & 0.1494 & 0.1484 & (0.2130, 0.3291, 1.4299, 1.1817, 17.1984) & $3.5218\times 10^{-6}$ & 0.9918 \\ \hline
2.4 & 0.10 & 0.1693 & 0.1659 & (0.2108, 0.2694, 0.6020, 1.2795, 21.1611) & $2.3385\times 10^{-6}$ & 0.9870 \\ \hline
2.4 & 0.15 & 0.1953 & 0.1915 & (0.1807, 0.1641, 0.2334, 1.4201, 22.5924) & $1.2139\times 10^{-6}$ & 0.9522 \\ \hline
2.4 & 0.20 & 0.2221 & 0.2178 & (0.1665, 0.1207, 0.1559, 1.5216, 22.8754) & $9.6766\times 10^{-7}$ & 0.9435 \\ \hline
4.7 & 0.05 & 0.4523 & 0.4201 & (0.4589, 0.3449, 1.0421, 1.5768, 35.9424) & $3.6264\times 10^{-5}$ & 0.9135 \\ \hline
4.7 & 0.10 & 0.5059 & 0.4769 & (0.4539, 0.2744, 0.3008, 1.7053, 54.1422) & $3.1442\times 10^{-5}$ & 0.9123 \\ \hline
16.5 & 0.22 & 2.0493 & 1.9328 & (0.6238, 0.1094, 0.0111, 4.4750, 105.3550) & $1.3212\times 10^{-6}$ & 0.9909 \\ \hline
23.6 & 0.22 & 3.3238 & 3.1952 & (0.7210, 0.1479, 0.0121, 7.4189, 65.6983) & $1.8010\times 10^{-6}$ & 0.9940 \\ \hline
\end{tabular}
}
\end{table}

\textbf{Direct Link Channel Model:}
In a direct link, the optical signal travels from the transmitter to the receiver along a single, unimpeded path. The intensity fluctuations of the signal caused by water turbulence in this path are modeled using a Gamma distribution. The PDF of the turbulence fading coefficient ($h_\alpha$) is defined as:
\[
f_{h_\alpha}\left(h_\alpha\right)=\frac{\alpha\left(\alpha h_\alpha\right)^{\alpha-1}}{\Gamma\left(\alpha\right)}\exp\left(-\alpha h_\alpha\right); \quad \alpha,\ h_\alpha>0
\]
Here, $\alpha$ is the turbulence parameter, which indicates the strength of the turbulence in the channel. This parameter depends on the scintillation index, $\sigma_l^2$, which is related to the signal wavelength, link distance, and water turbulence parameters such as the dissipation rate of turbulent kinetic energy ($\epsilon$) and the dissipation rate of mean-square temperature ($\chi_T$) (As explained in Appendix A). This model effectively describes a single propagation path.

\textbf{O-RIS Assisted Link Channel Model:}
In an O-RIS--assisted UWOC system, the communication path is divided into two cascaded segments:  
1) The link from the source to the O-RIS.  
2) The link from the O-RIS to the destination.  

Each of these two links is independently affected by water turbulence. The modeling of each sub-channel is done individually with a Gamma distribution.  

\vspace{5mm}
\textbf{PDF of the source--to--O-RIS link:}
{\scriptsize{
\[
f_{h_{\alpha_{_{s.r_m}}}}\left(h_{\alpha_{_{s.r_m}}}\right) =
\frac{\alpha \left(\alpha h_{\alpha_{_{s.r_m}}}\right)^{\alpha-1}}{\Gamma\left(\alpha\right)}
e^{-\alpha h_{\alpha_{_{s.r_m}}}}; \quad \alpha,\ h_{\alpha_{_{s.r_m}}}>0
\]
}}

\textbf{PDF of the O-RIS--to--destination link:}
{\scriptsize{
\[f_{h_{\alpha_{_{r_m.d}}}}\left(h_{\alpha_{_{r_m.d}}}\right)=\frac{\beta\left(\beta h_{\alpha_{_{r_m.d}}}\right)^{\beta-1}}{\Gamma\left(\beta\right)} e^{-\beta h_{\alpha_{_{r_m.d}}}}; \quad \beta,\ h_{\alpha_{_{r_m.d}}}>0 \]
}}

Here, $\alpha$ is the turbulence parameter for the source-to-O-RIS link and $\beta$ is for the O-RIS-to-destination link. Since the overall channel is a cascaded channel, and its total fading coefficient is the product of the fading coefficients of the two channels $h_{\alpha,m}=h_{\alpha_{s.r_m}}h_{\alpha_{r_m.d}},$ the overall channel distribution is a Gamma-Gamma distribution.  

To find the PDF of the product random variable $h_{\alpha,m}$, where $h_{\alpha_{s.r_m}}=x$, and $h_{\alpha_{r_m.d}}=y$; we use the general formula for the PDF of a product of two independent random variables:
\[
f_{h_{\alpha,m}}\left(z\right)=\int_{0}^{\infty}{f_{h_{\alpha_{s.r_m}}}\left(x\right)}\cdot f_{h_{\alpha_{r_m.d}}}\left(\frac{z}{x}\right)\cdot\frac{1}{x}\, dx
\]

By substituting the Gamma PDFs into this integral, we get:
{\scriptsize{\[ 
f_{h_{\alpha,m}}\left(z\right)=\int_{0}^{\infty}\left(\frac{\alpha^\alpha}{\Gamma\left(\alpha\right)}x^{\alpha-1}e^{-\alpha x}\right)\cdot\left(\frac{\beta^\beta}{\Gamma\left(\beta\right)}\left(\frac{z}{x}\right)^{\beta-1}e^{-\beta\frac{z}{x}}\right)\cdot\frac{1}{x}\, dx
\]}}

We simplify the integral by combining constant terms and powers of $x$:
\[
f_{h_{\alpha,m}}\left(z\right)=\frac{\alpha^\alpha\beta^\beta z^{\beta-1}}{\Gamma\left(\alpha\right)\Gamma\left(\beta\right)}\int_{0}^{\infty}x^{\alpha-\beta-1}\cdot e^{-\left(\alpha x+\frac{\beta z}{x}\right)}\, dx
\]

The resulting integral is in a standard form that can be solved using the modified Bessel function of the second kind $\left(K_v(\cdot)\right)$. The relevant integral identity is:
\[
\int_{0}^{\infty}{x^{v-1}e^{-ax-\frac{b}{x}}}\, dx=2\left(\frac{b}{a}\right)^{v/2}K_v\left(2\sqrt{ab}\right)
\]

By comparing our integral to this standard form, we can identify the following variable substitutions:  
$v=\alpha-\beta,\quad a=\alpha,\quad b=\beta z$ or by fixing $h_{\alpha_{s.r_m}}$ and using the change of variable $h_{\alpha_{r_m.d}}=h_{\alpha,m}/h_{\alpha_{s.r_m}}$, the conditional PDF can be obtained as:
\[
f\left(h_{\alpha,m}/h_{\alpha_{s.r_m}}\right)=\frac{\beta\left(\beta h_{\alpha,m}/h_{\alpha_{s.r_m}}\right)^{\beta-1}}{h_{\alpha_{s.r_m}}\Gamma\left(\beta\right)}\exp{\left(-\frac{\beta h_{\alpha,m}}{h_{\alpha_{s.r_m}}}\right)}
\]

In other words, by substituting these values, the integral evaluates to:
\[
\mathrm{Integral}=2\left(\frac{\beta z}{\alpha}\right)^{\left(\alpha-\beta\right)/2}K_{\alpha-\beta}\left(2\sqrt{\alpha\beta z}\right)
\]

Finally, by substituting the integral's result back into the PDF equation, we arrive at the final form of the Gamma-Gamma PDF:
\[
f_{h_{\alpha,m}}\left(h_{\alpha,m}\right)=\frac{2\left(\alpha\beta\right)^{\tfrac{\alpha+\beta}{2}}}{\Gamma\left(\alpha\right)\Gamma\left(\beta\right)}h_{\alpha,m}^{\tfrac{\alpha+\beta}{2}-1}K_{\alpha-\beta}\left(2\sqrt{\alpha\beta h_{\alpha,m}}\right)
\]

In this formula, $K_{\alpha-\beta}(\cdot)$ denotes the modified Bessel function of the second kind. The adopted two-stage model provides a more accurate characterization of the turbulence effects in indirect transmission channels employing an O-RIS. Specifically, this expression represents the PDF of the cascaded channel fading from the source to the destination via the $m$-th reflecting element. In the subsequent analysis, Eq.~(03.04.26.0009.01) in \cite{WolframResearch2025} was invoked to reformulate the Bessel function into a more tractable representation for further calculations. Furthermore, the property of the Meijer-$G$ function given in Eq.~(07.34.16.0001.01) of \cite{WolframResearch2025} was utilized to facilitate the analytical derivations.

{\small{
\begin{align*}
f_{h_\alpha}\left(h_\alpha\right) =&
\frac{A.B}{\Gamma(A).\Gamma(B).N}
\times (\frac{A.B}{N}.h_\alpha)^{\frac{A+B}{2}-1} \times \\ & G_{0,2}^{2,0} \left(
\frac{A.B}{N}. h_\alpha \ \Bigg| \begin{matrix}
\,- \\
\frac{A - B}{2}, \; \frac{B - A}{2}
\end{matrix}
\right)
\end{align*} }}

The PDF of the pointing error for the direct link is adopted from Eqs.~6 and 7 in~\cite{shetty2023performance}, and for the O-RIS case from Eqs.~5 and 12 in~\cite{wang2021performance}. 
To obtain the PDF of the instantaneous SNR with the aid of the PDF of $h$, we follow the procedure presented in \textit{Appendix A} of~\cite{naik2022evaluation}, which is expressed as follows:
\begin{align*}
f_h(h) &= \int_{0}^{A_0} f_{h_p}(h_p)\, f_{h|h_p}(h \,|\, h_p)\, \mathrm{d}h_p \\[6pt]
&= \int_{0}^{A_0} \frac{1}{h_p} f_{h_p}(h_p)\, f_\alpha\!\left(\frac{h}{h_p}\right)\, \mathrm{d}h_p ,
\end{align*}

\[
f_{h_{_D}}(h_{_D}) = \frac{\alpha \, \zeta^{2} \, (\alpha h_{_D})^{\zeta^{2}-1}}{A_0^{\zeta^{2}} \, \Gamma(\alpha)} 
\, G^{2,0}_{1,2}\!\left(\frac{\alpha h_{_D}}{A_0} \;\middle|\; \begin{array}{c} 1 \\ 0,\, \alpha - \zeta^{2} \end{array} \right)
\]

\begin{align*}
&f_{h_{_{O-RIS}}}(h_{_{O-RIS}}) = 
\frac{A. B. \xi^{2}}{N. \, \Gamma(A).\, \Gamma(B).\, A_0}
\,\\ & G^{3,0}_{1,3}\!\left(
\frac{A.B.h_{_{O-RIS}}}{N. A_0} \;\middle|\; 
\begin{array}{c}
\xi^{2} \\
\xi^{2}-1,\, A - 1,\, B - 1
\end{array}
\right).
\end{align*}

Then:
\begin{equation*}
f_{\gamma}(\gamma) = f_h \left( \sqrt{\tfrac{\gamma}{\bar{\gamma}}} \right) 
\left| \frac{dh}{d\gamma} \right|
= f_h \left( \sqrt{\tfrac{\gamma}{\bar{\gamma}}} \right) 
\left( \frac{1}{2\sqrt{\bar{\gamma}\gamma}} \right)
\end{equation*}

\begin{equation}
\begin{aligned}
    f_{\gamma_{_{O\!-\!RIS}}}(\gamma) &=
    \frac{A B \, \xi^2}{2N \, \Gamma(A)\Gamma(B)\, A_0 \sqrt{\bar{\gamma}\gamma}} \\
    &\quad \times G^{3,0}_{1,3} \left( 
    \frac{A B}{N A_0} \sqrt{\tfrac{\gamma}{\bar{\gamma}}} \;\middle|\;
    \begin{array}{c}
    \xi^2 \\[6pt]
    \xi^2 - 1, \, A - 1, \, B - 1
    \end{array}
    \right)
\end{aligned} \label{eq35}
\end{equation}

\begin{equation}
\begin{aligned}
f_{\gamma_{_D}}(\gamma) &= 
\left( \frac{\alpha}{A_0 \sqrt{\bar{\gamma}}} \right)^{\xi^2}
\frac{\xi^2}{2 \, \Gamma(\alpha)} \,
\gamma^{\tfrac{\xi^2}{2}-1} \,
\\& G^{2,0}_{1,2} \!\left(
\frac{\alpha}{A_0 \sqrt{\bar{\gamma}}} \, \gamma^{1/2}
\;\middle|\;
\begin{array}{c}
1 \\[6pt]
0, \; \alpha - \xi^2
\end{array}
\right)
\end{aligned} \label{eq36}
\end{equation}

The \textbf{outage probability} is calculated as follows:
\begin{equation}
OP = \int_{0}^{\gamma_{\text{th}}} f_{\gamma}(\gamma)\, d\gamma =F_{\gamma}(\gamma_{th}). \label{eq37}
\end{equation}

The closed-form expression of the outage probability is derived by substituting \eqref{eq36} and \eqref{eq35} into \eqref{eq37} and employing the integral identity of the Meijer-G function given in Eq.~(07.34.21.0084.01) of~\cite{WolframResearch2025}, as follows, respectively:

\vspace{-1mm} 
{\scriptsize
\begin{equation*}
\makebox[\linewidth][l]{%
  \scalebox{0.73}[1.1]{%
    \begin{minipage}{\linewidth}
    \setlength{\abovedisplayskip}{0pt}
    \setlength{\belowdisplayskip}{0pt}
    \begin{equation*}
    \begin{aligned}
    OP_D(\bar{\gamma}) = 
    &\left( \frac{\alpha}{A_0 \sqrt{\bar{\gamma}}} \right)^{\xi^2} 
    \times \frac{\xi^2 \cdot 2^{\alpha - \xi^2 - 2}}{\Gamma(\alpha) \cdot \sqrt{\pi}} 
    \times \gamma_{th}^{\frac{\xi^2}{2}} &\times G_{3,5}^{4,1} \left(
    \frac{\alpha^2 \gamma_{th}}{4 A_0^2 \bar{\gamma}} \,\Bigg|\, 
    \begin{array}{l}
    1 - \frac{\xi^2}{2},\ \frac{1}{2},\ 1 \\
    \frac{\alpha - \xi^2}{2},\ \frac{\alpha - \xi^2 + 1}{2},\ 0,\ \frac{1}{2},\ -\frac{\xi^2}{2}
    \end{array}
    \right)
    \end{aligned}
    \end{equation*}
    \end{minipage}
  }%
} \label{formula:2}
\end{equation*}
\noindent
\begin{equation*}
\makebox[\linewidth][l]{%
  \scalebox{0.72}[1.1]{%
    \begin{minipage}{\linewidth}
    \setlength{\abovedisplayskip}{0pt}
    \setlength{\belowdisplayskip}{0pt}
    \begin{equation*}
    \begin{aligned}
     OP_{O-RIS}(\bar{\gamma}) = \frac{2^{A+B-5} AB \xi^2}{\pi N \Gamma(A) \Gamma(B) A_0} 
    \left( \frac{\gamma_{th}}{\bar{\gamma}} \right)^{\frac{1}{2}} 
    \times G_{3,7}^{6,1} \left(
    \frac{A^2 B^2 \gamma_{th}}{16 N^2 A_0^2 \bar{\gamma}} \Bigg| 
    \begin{array}{l}
    \frac{1}{2}, \frac{\xi^2}{2}, \frac{\xi^2 + 1}{2} \\
    \frac{\xi^2 - 1}{2}, \frac{\xi^2}{2}, \frac{A - 1}{2}, \frac{B - 1}{2}, \frac{A }{2}, \frac{B }{2}, -\frac{1}{2}
    \end{array}
    \right)
    \end{aligned}
    \end{equation*}
    \end{minipage}
  }%
} \label{formula:3}
\end{equation*} }

For the \textbf{Bit Error Rate}, we derived the following closed-form expressions for both the direct link and the O-RIS--assisted link: 

\vspace{-5mm}
\begin{equation}
\textit{BER} = \frac{\delta}{2 \, \Gamma\left(p\right)} 
\sum_{k=1}^{\mathcal{K}} 
\left[ q_k^p \int_{0}^{\infty} e^{-q_k \gamma} \gamma^{p-1} F_\gamma(\gamma) \, d\gamma \right]. \label{eq:BER}
\end{equation}
The parameters $\delta$, $p$, $q_k$, and $\mathcal{K}$ depend on the modulation type, as detailed in Table~1 of~\cite{ramavath2020co}. 
The closed-form expression for the BER is obtained by using the Meijer--G integral given in Eq.~(07.34.21.0088.01) of~\cite{WolframResearch2025}. After substituting $F_{\gamma_D}(\gamma)$, and $F_{\gamma_{O\text{-}RIS}}(\gamma)$ into \eqref{eq:BER} and simplifying, we derive $\textit{BER}_D$ and $\textit{BER}_{O\text{-}RIS}$.

\vspace{-4mm}
\begin{table}[h!]
\centering
\caption{BER parameters for different modulation techniques~\cite{ramavath2020co}}
\vspace{-5pt}
\resizebox{\columnwidth}{!}{%
\begin{tabular}{|c|c|c|c|c|}
\hline
\textit{Modulation type} & \textbf{$p$} & \textbf{$\delta$} & \textbf{$q_k$} & \textbf{$\mathcal{K}$} \\ \hline
\textit{BFSK} & $1$ & $1$ & $0.5$ & $1$ \\ \hline
\textit{BPSK} & $0.5$ & $1$ & $1$ & $1$ \\ \hline
\textit{M-QAM} & $0.5$ & $\dfrac{4}{\log_2 M}\left(1-\dfrac{1}{\sqrt M}\right)$ 
& $\dfrac{3(2k-1)^2}{2(M-1)}$ & $\max \left(\lfloor \frac{\sqrt{M}}{2} \rfloor, 1 \right)$ \\ \hline
\textit{M-PSK} & $0.5$ & $\dfrac{2}{\max\left(\log_2 M , 2\right)}$ 
& $\sin^2\!\left(\dfrac{(2k-1)\pi}{M}\right)$ & $\max(\frac{M}{4}, 1)$ \\ \hline
\end{tabular}%
}
\end{table}
{
\scriptsize 
\begin{equation*}
\makebox[\linewidth][l]{%
  \scalebox{0.72}[1.2]{%
$\displaystyle \textit{BER}_D = 
\frac{\delta}{2\Gamma(p)} \sum\limits_{k=1}^{\mathcal{K}}  
 \frac{\zeta^2 \cdot 2^{\alpha - \zeta^2 - 2}}{\sqrt{\pi} \Gamma(\alpha)}
\left( \frac{\alpha}{A_0 \sqrt{q_k \bar{\gamma}}} \right)^{\zeta^2} \cdot
G_{4,5}^{4,2} \left(
\frac{\alpha^2}{4 A_0^2 q_k \bar{\gamma}} \left|
\begin{array}{c}
1 - p - \frac{\zeta^2}{2},\ 1 - \frac{\zeta^2}{2},\ \frac{1}{2},\ 1 \\
\frac{\alpha - \zeta^2}{2},\ \frac{\alpha - \zeta^2 + 1}{2},\ 0,\ \frac{1}{2},\ -\frac{\zeta^2}{2}
\end{array}
\right.
\right)$}}
\end{equation*}
\begin{equation*}
\makebox[\linewidth][l]{%
  \scalebox{0.67}[1.2]{%
$\displaystyle
\textit{BER}_{\mathrm{O\text{-}RIS}} =
\frac{\delta}{2 \Gamma(p)} 
\sum_{k=1}^{\mathcal{K}} 
\frac{2^{A + B - 5} \cdot AB \cdot \zeta^2}{\pi N A_0 \Gamma(A) \Gamma(B) \sqrt{q_{_k} \bar{\gamma}}}
\cdot
G_{4,7}^{6,2} \left(
\frac{A^2 B^2}{16 q_k N^2 A_0^2 \bar{\gamma}}
\left|
\begin{array}{c}
\frac{1}{2},\, \frac{1}{2} - p,\, \frac{\zeta^2}{2},\, \frac{\zeta^2 + 1}{2} \\
\frac{\xi^2 - 1}{2},\, \frac{\xi^2}{2},\, \frac{A - 1}{2},\, \frac{B - 1}{2},\, \frac{A}{2},\, \frac{B}{2},\, -\frac{1}{2}
\end{array}
\right.
\right)$ }}
\end{equation*}
}

As reported in \cite{naik2022evaluation, R-Zedini}, the \textbf{channel capacity} under HD and IM/DD, in the absence of channel state information (CSI) at the transmitter, is expressed as:

\begin{equation}
    C = \int_{0}^{\infty} \log_{2}\!\bigl(1+\tau \gamma\bigr)\, f_{\gamma}(\gamma)\, d\gamma ,
\end{equation}

The closed-form expression of the channel capacity is obtained by utilizing the integration of the Meijer-G function presented in Eq. (01.04.26.0003.01) in \cite{WolframResearch2025} and the substitution of \eqref{eq36} and \eqref{eq35}. Afterward, we simplify the expressions and derive the closed-form capacities, denoted as $C_{_D}(\bar{\gamma})$ and $C_{_{O-RIS}}(\bar{\gamma})$, respectively:

{\scriptsize
\begin{equation*}
\makebox[\linewidth][l]{%
  \scalebox{0.72}[1.2]{%
    \begin{minipage}{\linewidth}
    \setlength{\abovedisplayskip}{0pt}
    \setlength{\belowdisplayskip}{0pt}
    \begin{equation*}
    \begin{aligned}
    \hspace{0.5cm}C_D(\bar{\gamma}) = \left( \frac{\alpha}{A_0 \sqrt{\tau \bar{\gamma}}} \right)^{\zeta^2}
    \frac{\zeta^2 2^{(\alpha - \zeta^2 - 2)}}{\sqrt{\pi} \Gamma(\alpha) \ln(2)} 
   \times G_{4,6}^{6,1} \left( 
   \frac{\alpha^2}{4 A_0^2 \tau \bar{\gamma}} 
   \Bigg| 
   \begin{array}{c}
   \frac{-\zeta^2}{2}, 1 - \frac{\zeta^2}{2}, \frac{1}{2}, 1 \\
   \frac{\alpha - \zeta^2}{2}, \frac{\alpha - \zeta^2 + 1}{2}, 0, \frac{1}{2}, -\frac{\zeta^2}{2}, -\frac{\zeta^2}{2}
   \end{array}
    \right)
    \end{aligned}
    \end{equation*} \label{formula:9}
    \end{minipage}
  }%
}
\end{equation*}
\noindent 
\begin{equation*}
\makebox[\linewidth][l]{%
  \scalebox{0.695}[1.2]{%
    \begin{minipage}{\linewidth}
    \setlength{\abovedisplayskip}{0pt}
    \setlength{\belowdisplayskip}{0pt}
    \begin{equation*}
    \begin{aligned}
C_{O-RIS}(\bar{\gamma}) = \frac{2^{A + B - 5} AB \zeta^2}{ \pi N \ln(2) \Gamma(A) \Gamma(B) A_0 \sqrt{\tau \bar{\gamma}}} \times G_{4,8}^{8,1} \left( 
\frac{A^2 B^2}{16 N^2 A_0^2 \tau \bar{\gamma}} 
\Bigg| 
\begin{array}{c}
\frac{-1}{2}, \frac{1}{2}, \frac{\zeta^2}{2}, \frac{\zeta^2 + 1}{2} \\
\frac{\xi^2 - 1}{2}, \frac{\xi^2}{2}, \frac{A - 1}{2}, \frac{B - 1}{2}, \frac{A }{2}, \frac{B }{2}, -\frac{1}{2}, -\frac{1}{2}
\end{array}
\right)
    \end{aligned}
    \end{equation*}
    \end{minipage}
  }%
} \label{formula:10}
\end{equation*} }
Here, $\tau$ represents the detection type, where $\tau = 1$ for HD and $\tau = \frac{e}{2\pi}$ for IM/DD \cite{R-Zedini}.

\vspace{2mm}
For the \textbf{Diversity Order} and \textbf{Speed of Convergence}, following \cite{R41, R42} and the approach in \cite{zhang2024performance}, we adopt the derivative of the Meijer-$G$ function (Eq.~(07.34.20.0002.01) in \cite{WolframResearch2025}) combined with the differentiation chain rule. By subsequently applying Eq.~(07.34.06.0006.01) in \cite{WolframResearch2025}, which provides the series expansion of the Meijer-$G$ function around $z \to 0$, and by exploiting the recursive property of the Gamma function, $\Gamma(x+1)=x\Gamma(x)$, the Meijer-$G$ function can be expressed in a summation form with coefficients involving Gamma terms. Accordingly, the Diversity Order is defined as:
\begin{equation*}
\mathrm{DO}(\bar{\gamma}) = - \frac{\partial \ln  OP(\bar{\gamma})}{\partial \ln \bar{\gamma}} .
\label{eq:DO_def}
\end{equation*}
Based on the derivations, the DO can be expressed as
{ \scriptsize { \begin{equation}
\begin{aligned}
\mathrm{DO}(\bar{\gamma}) &= \tfrac{1}{2} + 
\frac{ G^{6,2}_{4,8}\!\left( 
\frac{A^2 B^2 \gamma_{\text{th}}}{16 N^2 A_0^2 \bar{\gamma}}
\;\middle|\;
\begin{array}{c}
0,\; \tfrac{1}{2},\; \tfrac{\xi^2}{2},\; \tfrac{\xi^2+1}{2} \\
\frac{\xi^2 - 1}{2}, \frac{\xi^2}{2}, \frac{A - 1}{2}, \frac{B - 1}{2}, \frac{A }{2}, \frac{B }{2},\; 1,\; -\tfrac{1}{2}
\end{array}
\right) }
{ G^{6,1}_{3,7}\!\left( 
\frac{A^2 B^2 \gamma_{\text{th}}}{16 N^2 A_0^2 \bar{\gamma}}
\;\middle|\;
\begin{array}{c}
\tfrac{1}{2},\; \tfrac{\xi^2}{2},\; \tfrac{\xi^2+1}{2} \\
\frac{\xi^2 - 1}{2}, \frac{\xi^2}{2}, \frac{A - 1}{2}, \frac{B - 1}{2}, \frac{A }{2}, \frac{B }{2},\; -\tfrac{1}{2}
\end{array}
\right) } .
\end{aligned}
\label{eq:DO_final}
\end{equation} }}

The \textbf{asymptotic diversity order} characterizes the rate at which the outage probability decreases in the high-SNR regime \cite{zhang2024performance}. It is defined in the formula as follows:
\begin{equation*}
\text{ADO} = \underset{{\bar{\gamma} \to \infty} }{\text{DO}(\bar{\gamma})}= - \lim_{\bar{\gamma} \to \infty} 
\frac{\partial \ln \left[ OP(\bar{\gamma}) \right]}{\partial \ln \bar{\gamma}} .
\end{equation*}

By further analyzing the outage probability expression, the ADO can be simplified as
\begin{align}
\text{ADO} 
&= \frac{1}{2} \min \left(A, \xi^{2},B \right) .
\end{align}
This result reveals that the ADO depends on the weakest parameter among the channel and system characteristics. 
Hence, the overall ADO is determined by half of the minimum of these parameters. This means that the limiting factor (the smallest among channel fading or misalignment) dominates the achievable diversity order in the asymptotic regime.

\vspace{5mm}
The  \textbf{Speed Of Convergence} measures how fast the instantaneous diversity order $\text{DO}(\bar{\gamma})$ approaches the asymptotic value ADO. It is defined as:
\begin{equation}
\text{SOC}(\bar{\gamma}) 
= - \frac{1}{\text{ADO}} 
\frac{\partial^{2} \ln \text{OP}(\bar{\gamma})}{\partial (\ln \bar{\gamma})^{2}}
= \frac{1}{\text{ADO}} 
\frac{\partial \text{DO}(\bar{\gamma})}{\partial \ln \bar{\gamma}} .
\end{equation}
\noindent
Eq. \eqref{eq43} indicates that SOC is normalized by the ADO and depends on the second derivative of the logarithm of the OP. A large value of SOC implies that the actual diversity order converges faster to its asymptotic value, which in turn validates the use of asymptotic analysis even at moderate SNR values. A closed-form expression for SOC, obtained via Meijer-G function representations, is given as \cite{zhang2024performance}:

$\hspace{3.2cm}\text{SOC}(\bar{\gamma})
= $ 
{\scriptsize{\begin{equation}
\begin{aligned}
&\Bigg\{ 2\times
\left[ 
G^{6,2}_{4,8} \!\left( 
\frac{A^{2}.B^{2}.\gamma_{th}}{16.N^{2}.A_0^{2}.\bar{\gamma}}
\ \Bigg| \ 
\begin{array}{c}
0, \tfrac{1}{2}, \tfrac{\xi^{2}}{2}, \tfrac{\xi^{2}+1}{2} \\
\frac{\xi^2 - 1}{2}, \frac{\xi^2}{2}, \frac{A - 1}{2}, \frac{B - 1}{2}, \frac{A }{2}, \frac{B }{2}, 1, -\tfrac{1}{2}
\end{array}
\right) 
\right]^{2} \\[6pt]
&\quad - 2 \times G^{6,3}_{5,9} \!\left(
\frac{A^{2}.B^{2}.\gamma_{th}}{16.N^{2}.A_0^{2}.\bar{\gamma}}
\ \Bigg|\ 
\begin{array}{c}
0, 0, \tfrac{1}{2}, \tfrac{\xi^{2}}{2}, \tfrac{\xi^{2}+1}{2} \\
\frac{\xi^2 - 1}{2}, \frac{\xi^2}{2}, \frac{A - 1}{2}, \frac{B - 1}{2}, \frac{A }{2}, \frac{B }{2}, 1, 1, -\tfrac{1}{2}
\end{array}
\right) \\[6pt]
&\quad \times G^{6,1}_{3,7} \!\left(
\frac{A^{2}.B^{2}.\gamma_{th}}{16.N^{2}.A_0^{2}.\bar{\gamma}}
\ \Bigg|\ 
\begin{array}{c}
\tfrac{1}{2}, \tfrac{\xi^{2}}{2}, \tfrac{\xi^{2}+1}{2} \\
\frac{\xi^2 - 1}{2}, \frac{\xi^2}{2}, \frac{A - 1}{2}, \frac{B - 1}{2}, \frac{A }{2}, \frac{B }{2}, -\tfrac{1}{2}
\end{array}
\right) \Bigg\} \\[6pt]
&\quad \hspace{2cm} \Bigg/ \Bigg\{
\min\!\left(A,\xi^{2}, B\right) \times \\ &
\hspace{5mm}\Bigg[
G^{6,1}_{3,7} \!\left(
\frac{A^{2}.B^{2}.\gamma_{th}}{16.N^{2}.A_0^{2}.\bar{\gamma}}
\ \Bigg|\ 
\begin{array}{c}
\tfrac{1}{2}, \tfrac{\xi^{2}}{2}, \tfrac{\xi^{2}+1}{2} \\
\frac{\xi^2 - 1}{2}, \frac{\xi^2}{2}, \frac{A - 1}{2}, \frac{B - 1}{2}, \frac{A }{2}, \frac{B }{2}, -\tfrac{1}{2}
\end{array}
\right)
\Bigg]^{2} \Bigg\}.
\end{aligned} \label{eq43}
\end{equation} }}

\noindent
Although the final expression involves multiple Meijer-G functions, it provides a rigorous closed-form description of how fast the diversity order converges to its asymptotic value under the considered fading and misalignment conditions.

\section{Appendix E}
\subsection{Proposed UWOC-WDM-PON Architecture}

In the proposed architecture, the uplink data flow begins in the UWOC, which provides an effective bandwidth of approximately $100\,\text{MHz}$. At the UOAP, a practical aggregate spectral efficiency of 
$150~\text{bps/Hz}$ is achieved, corresponding to a raw data rate 
of about $15~\text{Gbps}$, which results from the aggregation of 
five transceivers. The received signals are processed through demodulation, followed by packetization or framing operations when required. Subsequently, the signals are transmitted over standard single-mode fiber (SMF, compliant with ITU-T G.652.D) to the ONU. 
At the ONU, the 25GS-PON upstream transceiver operates on a dedicated wavelength of $\lambda_{\mathrm{UL}} = 1286 \pm 2$~nm (UW3 coexistence option), explicitly defined in the 25GS-PON MSA to ensure coexistence with GPON and XGS-PON. Each upstream channel provides a line rate of 24.8832~Gbps. With LDPC FEC, the net payload reaches approximately 21.17~Gbps, which comfortably accommodates the 15~Gbps raw UWOC throughput.

Upstream signals from multiple ONUs are aggregated in the Optical Distribution Network (ODN) through an Arrayed Waveguide Grating (AWG) or Mux/Demux. Unlike TDM-PON, which employs a shared single wavelength, WDM-PON assigns a dedicated wavelength to each ONU, enabling true rate aggregation. The resulting O-band channels are transported over the \textbf{feeder fiber} toward the metro central office. The feeder fiber is a standard ITU-T G.652.D single-mode fiber with an O-band attenuation of approximately $0.35~\text{dB/km}$. For an $80~\text{km}$ distance, the resulting loss remains within the optical power budget of long-reach PON classes (N1/N2) and is compensated at the ONU using high-sensitivity receivers (e.g., APD/TIA) and sufficient launch power, eliminating the need for optical amplification. Hence, fiber attenuation is predictable and non-critical in the system design. In contrast, UWOC channels suffer from severe and environment-dependent attenuation due to absorption and scattering, often exceeding fiber losses by orders of magnitude. Therefore, this study neglects fiber attenuation and focuses on the UWOC segment, which represents the dominant performance-limiting factor of the hybrid system.

For longer-reach applications, O-band channels can be mapped via transponders or OEO conversion to C-band DWDM channels (ITU-T G.694.1 compliant, 100~GHz spacing). This enables seamless integration with metro/backhaul networks while maintaining signal quality. At the OLT, these upstream channels are demultiplexed and terminated at their corresponding ports, delivering approximately 15~Gbps net per branch and about 240~Gbps total upstream capacity with 16 branches. For the downlink, data originates at the OLT in the metro central office. The 25GS-PON downstream ports generate service traffic on the preferred downstream wavelength $\lambda_{\mathrm{DL}} = 1358$~nm in the O-band.

Each downstream channel provides a line rate of 24.8832~Gbps, with a net payload of $\approx 21.17$~Gbps after LDPC FEC. Downstream wavelengths can also be transported over a DWDM metro backbone if required, and are subsequently demultiplexed in the ODN using an AWG to direct each wavelength to its corresponding ONU. The signals then travel over SMF G.652.D to each ONU and are forwarded to the UOAPs, which convert the optical downstream signal into the UWOC domain. Thus, each branch supports $\approx 15$~ Gbps of net downstream throughput, and with 16 branches, the aggregate downstream capacity reaches $\approx 240$~Gbps.

The system supports bidirectional traffic over dedicated O-band wavelengths: uplink at $1286 \pm 2$~nm (UW3 coexistence option) and downlink at 1358~nm (25GS-PON preferred downstream). Each channel provides a line rate of 24.8832~Gbps and a net payload of $\approx 21.17$~Gbps after LDPC FEC, which is sufficient to carry the 15~Gbps raw UWOC throughput per branch. With 16 independent WDM-PON channels, the aggregate capacity reaches $\approx 240$~Gbps in the uplink and $\approx 240$~Gbps in the downlink, resulting in a total bidirectional capacity of nearly 480~Gbps.
Therefore, the total capacity of 480~Gbps in this architecture is obtained by aggregating 16 independent 25GS-PON channels (each with a line rate of approximately 25~Gbps and a net payload of approximately 21~Gbps ) in each direction. In other words, the WDM-PON access network provides an aggregate of approximately 240~Gbps in the \textit{uplink} and 240~Gbps in the \textit{downlink} over the \textit{feeder fiber} (480~Gbps full-duplex). This capacity results from the sum of multiple channels and does not correspond to a single 480~Gbps channel.

This architecture is compliant with 25GS-PON MSA (IEEE~802.3ca, ITU-T~G.9807.1), ITU-T~G.652.D, and ITU-T~G.694.1 standards. 
The design ensures sufficient optical budget, minimizes latency, guarantees dedicated bandwidth per branch, and provides scalability for future expansion with additional wavelengths or higher data rates.

\subsection{Analysis of PON Performance based on \textbf{Delay and Throughput}}

Considering the negligible level of inter-fiber noise and loss when integrating with UOAP and UWOC, these effects have been disregarded. Accordingly, two important evaluation parameters—Delay and Throughput—are taken into account for the backhaul segment~\cite{liu2024ultraLowDelayPON, Haastrup2023}.

Core Network Design Parameters:

The upstream bandwidth of the feeder fiber is $R_{\text{upstream},(POS\!-\!UOAP)} = 15~\mathrm{Gbps}$, whereas the total aggregate upstream bandwidth from the UOAPs to the OLT is $R_{\text{upstream},(OLT\!-\!POS)} = 240~\mathrm{Gbps}$. The fiber length is $L = 80~\mathrm{km}$, with a propagation delay of 
$T_{\text{prop},km} = 5\times10^{-6}~\mathrm{s/km}$. 
A fixed FEC delay of $T_{\text{FEC}} = 0.25~\mathrm{ms}$ and a fixed processing delay of 
$T_{\text{proc}} = 0.2~\mathrm{ms}$ are considered. 
The number of UOAPs (ONUs) is $\mathfrak{T}\in\{2,\dots,16\}$, 
and the base delay is set to $T_{\text{base}} = 0.4~\mathrm{ms}$. 
The report request time and grant send time are 
$T_{\text{report}} = 0.4~\mu\mathrm{s}$ and 
$T_{\text{grant}} = 0.4~\mu\mathrm{s}$, respectively. 
In addition, the guard time is $T_{\text{guard}} = 100~\mathrm{ns}$, 
and the service time is $T_{\text{service}} = 100~\mu\mathrm{s}$. 
The processing coefficient is defined as 
$k = 0.005~\mathrm{ms}/\mathfrak{T}^2$, 
with a power-law exponent of $\alpha_{\text{p.l.}} = 2.5$.

\hspace{3mm}
\noindent\subsubsection{Total Delay Modeling} The total end-to-end delay ($T_{total}$) for each UOAP is the sum of several components:
\begin{equation}
    T_{total}=T_{DBA}+T_{FEC}+T_{proc}+T_{queue}+T_{prop}+T_{retrans}
\end{equation}
\hspace{5mm}
Decomposition of Delay Factors:

\textbf{Propagation Delay ($T_{prop}$)}: The time a signal needs to travel the fiber length.
\[T_{prop}=L\times T_{prop,km} \]

\textbf{DBA Delay ($T_{DBA}$):} The total DBA delay is the sum of a constant base delay, a polling-related delay, and a processing-related delay. The general formula is:
\[T_{DBA}=T_{base}+T_{polling}+T_{proc} \quad,\]
$T_{base}$ is a constant base delay representing the minimum OLT processing overhead.
$T_{polling}$ is the delay component caused by the polling mechanism, which scales linearly with the number of ONUs $(\mathfrak{T})$. The formula is given by:
\[T_{polling}=\frac{1}{2}\times \mathfrak{T}\times\left(T_{report}+T_{grant}+T_{guard}+T_{service}\right)\]

$T_{proc}$ models the computational complexity of the DBA algorithm and can be represented by a power-law relationship: \[T_{proc}=k\times \mathfrak{T}^{\alpha_{p.l.}}\]

\textbf{Queuing Delay ($T_{queue}$)}: This is an empirical power-law model for PON queuing behavior. The formula is:

{\small$T_{delay\_queue}=T_{base\_queue}+K_{load\_factor}\times\left(\rho_{load}\right)^{\alpha_{p.l.}}+\zeta_{noise}$}

\noindent In these formulas, $\zeta_{noise}$ is a random value following a normal distribution.

\textbf{Retransmission Delay ($T_{retrans}$)}: Delay due to packet loss and the need for retransmission.
\[T_{retrans}=\mathrm{packet\_loss\_rate}\times0.4\times{10}^{-3}\]

\subsubsection{Throughput Modeling} The network throughput is analyzed for both ideal and realistic scenarios. 

Ideal Throughput ($T_{ideal}$): This throughput is calculated without considering any network overhead.  The network capacity is divided among users, considering network efficiency ($\eta$) and the physical link's maximum capacity ($C_{max}$).
\[TH_{\mathrm{ideal}}\left(\mathfrak{T}\right)={min}\left(\frac{\eta\cdot R_{\mathrm{upstream}}}{\mathfrak{T}},\;C_{max}\right)\]
Where $R_{\mathrm{upstream}}$ is the Total upstream network capacity and $\mathfrak{T}$ is the Number of UOAPs (User-side Optical Access Points also,  $\eta$ represents Network efficiency and  $C_{max}=15\mathrm{\ Gbps\ }$is the maximum capacity of the UOAP to the passive optical splitter (POS) link.

\textbf{Realistic Throughput ($TH_{real}$)}: To compute realistic throughput, we first model the traffic demand ($D_i$) using a Zipf distribution and then allocate resources based on this demand and network constraints.

\textbf{Zipf Demand Model}: The demand of each UOAP ($D_i$) is calculated based on its rank ($rank_i$) in the Zipf distribution~\cite{7493685}:
\[D_i=\left(\frac{\frac{1}{rank_i^\alpha}}{\sum_{j=1}^{\mathfrak{T}}\frac{1}{rank_j^\alpha}}\right)\times\left(1.6\times R_{upstream}\right)\]
Here, $\alpha=\ 1$ is the Zipf exponent, and 1.6 is an overload factor to simulate high-traffic conditions.

\textbf{Allocated Capacity per UOAP:} Initially, the raw allocated capacity is computed based on demand and system efficiency.
\[A_i^{\mathrm{raw}}=\frac{D_i}{\sum_{j=1}^{\mathfrak{T}}D_j}\cdot\left(\eta\cdot R_{\mathrm{upstream}}\right)\]
Next, the theoretical capacity limit for each UOAP is calculated based on the number of UOAPs, i.e.:  
\[C_{\mathrm{share}}=\frac{\eta\cdot R_{\mathrm{upstream}}}{\mathfrak{T}} \]. 
Then, the final allowed capacity limit for each UOAP is: 
\[C_i={min}\left(C_{max},\;C_{\mathrm{share}}\right)\]. 
Finally, the actual allocated capacity (without considering packet loss) is:
\[A_i={min}\left(D_i\;,{min}\left(A_i^{\mathrm{raw}},\;C_i\right)\right)\]

\textbf{Overload Ratio:} The network's overload ratio is the ratio of total network load to its effective capacity.
\[\gamma={max}\left(\frac{\sum_{i=1}^{\mathfrak{T}}D_i}{\eta\cdot R_{\mathrm{upstream}}},\;1\right)\]

\textbf{Packet Loss Rate:} The packet loss rate is modeled based on the overload ratio.
\[P_{\mathrm{loss}}={min}\left(0.25,\;0.005\cdot\gamma^{2.5}\right)\]

\textbf{Effective Throughput:} Considering packet loss, the effective throughput for each UOAP is:
\[TH_{\mathrm{real}}\left(i\right)=A_i\cdot\left(1-P_{\mathrm{loss}}\right)\]
Ultimately, the expression for $T_{\mathrm{real}}\left(i,\mathfrak{T}\right)$  can be written as:
\makebox[\linewidth][l]{%
    \scalebox{0.66}[.9]{%
        \begin{minipage}{\linewidth}
        \begin{equation}
        \begin{aligned}
        TH_{\mathrm{real}}(i,\mathfrak{T}) &= \min \left( D_i,\; \min \left( \frac{D_i}{\sum_{j=1}^{\mathfrak{T}} D_j} \cdot (\eta \cdot R_{\mathrm{upstream}}), \min \left( C_{max}, \; \frac{\eta \cdot R_{\mathrm{upstream}}}{\mathfrak{T}} \right) \right) \right) \\ 
        &\quad \cdot \left( 1 - \min \left( 0.25,\;  0.005 \cdot  \left( \max \left( \frac{\sum_{k=1}^{\mathfrak{T}} D_k}{\eta \cdot R_{\mathrm{upstream}}},\; 1 \right) \right)^{2.5} \right) \right) 
        \end{aligned}
        \end{equation}
        \end{minipage}%
    }%
}
This value is then converted to Gbps for the final result. 

\subsection{Hard Switching and Optimal Link Selection}

Furthermore, the overall system capacity and the switching considerations for the entire system are taken into account. A similar approach for two i.i.d. links is considered in~\cite{MOHSAN2023100697}. For link selection between the direct and O-RIS--assisted paths in each transceiver, a hard switching mechanism is employed, which is modeled as:

{\small{
\begin{equation}
C_{_\mathrm{selected}}=\left\{\begin{matrix}C_D\left(\bar{\gamma}\right),&\mathrm{if\ }C_D\left(\bar{\gamma}\right)\geq C_{\mathrm{th}},\\C_{O-RIS}\left(\bar{\gamma}\right),&\mathrm{if\ }C_D\left(\bar{\gamma}\right)<C_{\mathrm{th}},\\\max{\{}C_D\left(\bar{\gamma}\right),C_{O-RIS}\left(\bar{\gamma}\right)\},&\text{\scriptsize{optimal\ selection\ case}}.\\\end{matrix}\right.
\end{equation}}}

Here, $C_{\mathrm{th}}$ is the capacity threshold, determined according to the required quality-of-service (QoS) level. This mechanism ensures that when the direct link suffers from turbulence, blockage, or severe SNR degradation, the O-RIS--assisted link is rapidly activated to maintain system throughput. 

The simulation results presented in Fig.4(b) show that for a configuration with two UOAPs, each equipped with three direct links and two O-RIS--assisted links, the hard switching mechanism significantly improves total capacity in low-SNR regimes compared to using direct links only. Increasing the number of O-RIS elements (N) yields a noticeable capacity gain, particularly in medium-to-high SNR regions. However, when the PON backhaul becomes the bottleneck, the total capacity cannot exceed the upper bound $C_{\mathrm{POS-UOAP}}$, regardless of improvements in the UWOC link performance. In the optimal selection scenario, where the link with the highest instantaneous capacity is chosen at each SNR value, the system achieves maximum resource utilization; however, this approach increases control complexity and requires real-time link capacity estimation. Therefore, an effective design must balance UWOC link performance, the number of O-RIS elements, and backhaul bandwidth allocation policies to achieve both high capacity and stable operation.

\section{Appendix F}
\subsubsection{Impact of the Number of O-RIS Elements on the Statistical Characteristics of the UWOC Link Under Gamma–Gamma Turbulence}

In our model, the turbulence-induced fading of each individual O-RIS link, denoted by $h_{\alpha,m}$, is modeled as an independent and identically distributed (i.i.d.) Gamma–Gamma random variable with parameters $\left(\alpha,\beta,\Omega\right)$. When the total O-RIS gain is expressed as:
\begin{equation}
    h_\alpha=\sum_{m=1}^{N}h_{\alpha,m},
\end{equation}
Its PDF does not admit a simple closed form due to the convolution of $N$ G–G distributions. To obtain a tractable analytical expression, we follow the well-known moment-matching approximation method reported in \cite{8438298} (and also widely used in optical wireless literature). Specifically, \cite{8438298} shows that the sum of $N$ i.i.d. G–G random variables can be accurately approximated by another G–G variable with parameters:
\begin{equation}
    A=c\alpha,\quad  B=c\beta,\quad  \bar{\Omega}=c\Omega,
\end{equation}
where the correction factor $c$ is given by:
\begin{equation}
    c=N\left(1+\frac{2}{N}\sum_{n=1}^{N}\sum_{m=n+1}^{N}\rho_{nm}\right)^{-1},
\end{equation}
and $\rho_{nm}$ is the correlation coefficient between $h_{\alpha,n}$ and $h_{\alpha,m}$. In the i.i.d. independent case considered here ($\rho_{nm}=0$), this reduces to $c = N$, yielding:
\begin{equation}
    A=N\alpha,\quad  B=N\beta,\quad  \bar{\Omega}=N\Omega.
\end{equation}

This approximation preserves the first and second moments of the sum, ensuring that the mean and variance match those of the exact sum distribution. It has been shown in \cite{8438298} and other works in the optical wireless literature to be highly accurate even for small and large $N$, which supports its adoption in our analysis.

\subsubsection*{Proof of the Proposition in Appendix A of \cite{8438298}}
Suppose that the sum of $N$ identically but not necessarily independently distributed G–G random variables (RVs) 
\[
X_i\sim\Gamma\Gamma\left(\alpha,\beta,\Omega\right)\quad \left(i=1,\ldots,N\right)
\]
is approximated by a G–G RV 
$
Y\sim\Gamma\Gamma\left(\bar{\alpha}_s,\bar{\beta}_s,\Omega_s\right),
$
i.e., $Y\approx\sum_{i=1}^{N}X_i$. The first and second moments of $\sum_{i=1}^{N}X_i$ are given by 
\begin{equation}
    \mathbb{E}\left[\sum_{i=1}^{N}X_i\right]=\sum_{i=1}^{N}\mathbb{E}\left[X_i\right]=N\Omega,
\end{equation}
and
{ \small
\begin{align}
    &\mathbb{E}\!\left[\left(\sum_{i=1}^{N}X_i\right)^2\right] 
    = \sum_{i=1}^{N}\mathbb{E}\!\left[X_i^2\right] 
    + 2\sum_{i=1}^{N}\sum_{j=i+1}^{N}\mathbb{E}\!\left[X_iX_j\right] \nonumber= \sum_{i=1}^{N} \\
    & \mathbb{E}\!\left[X_i^2\right]  + 2\sum_{i=1}^{N}\sum_{j=i+1}^{N}\Bigg(
        \mathbb{E}\!\left[X_i\right]\mathbb{E}\!\left[X_j\right]
        + \rho_{ij}\sqrt{\mathrm{Var}\!\left(X_i\right)\,\mathrm{Var}\!\left(X_j\right)}
      \Bigg) \nonumber\\
    &= \sum_{i=1}^{N}\mathbb{E}\!\left[X_i^2\right] 
    + 2\sum_{i=1}^{N}\sum_{j=i+1}^{N}\left(\Omega^2+\rho_{ij}\sigma^2\right).
\end{align}
}

The variance of $\sum_{i=1}^{N}X_i$ is then given by
\begin{align}
    \mathrm{Var}\left(Y\right) &= \mathbb{E}\left[\left(\sum_{i=1}^{N}X_i\right)^2\right]-\left(\mathbb{E}\left[\sum_{i=1}^{M}X_i\right]\right)^2 \nonumber\\
    &= \sum_{i=1}^{N}\mathrm{Var}\left(X_i\right)+2\sum_{i=1}^{N}\sum_{j=i+1}^{N}\rho_{ij}\sigma^2 \nonumber\\
    &= \left(\frac{1}{\alpha}+\frac{1}{\beta}+\frac{1}{\alpha\beta}\right)\left(N+2\sum_{i=1}^{N}\sum_{j=i+1}^{N}\rho_{ij}\right)\Omega^2.
\end{align}

On the other hand, the mean and variance of $Y$ are written as
$
    \mathbb{E}\left[Y\right]=\Omega_s,
$
, and $
    \mathrm{Var}\left(Y\right)=\left(\frac{1}{\bar{\alpha}_s}+\frac{1}{\bar{\beta}_s}+\frac{1}{\bar{\alpha}_s\bar{\beta}_s}\right)\bar{\Omega}_s^2.
$

Matching the mean and variance of $Y$ with that of $\sum_{i=1}^{N}X_i$ yields $\Omega_s=N$, and
$
    N^2\left(\frac{1}{\bar{\alpha}_s}+\frac{1}{\bar{\beta}_s}+\frac{1}{\bar{\alpha}_s\bar{\beta}_s}\right)=\left(\frac{1}{\alpha}+\frac{1}{\beta}+\frac{1}{\alpha\beta}\right)\left(N+2\sum_{i=1}^{N}\sum_{j=i+1}^{N}\rho_{ij}\right).
$

To solve for $\bar{\alpha}_s$ and $\bar{\beta}_s$, it is required to have one more equation, which can be obtained from matching higher moments. However, this leads to complex expressions. For simplicity, an approximate solution is obtained as:
\begin{equation}
    \bar{\alpha}_s=N\left(1+\frac{2}{N}\sum_{i=1}^{N}\sum_{j=i+1}^{N}\rho_{ij}\right)^{-1}\alpha,
\end{equation}
\begin{equation}
    \bar{\beta}_s=N\left(1+\frac{2}{N}\sum_{i=1}^{N}\sum_{j=i+1}^{N}\rho_{ij}\right)^{-1}\beta.
\end{equation}

\subsubsection{Impact of the Number of O-RIS Elements on the Statistical Characteristics of the UWOC Link Under mEGG Turbulence} 

As described in Section C in \cite{10413214}, the total reflected gain from the O-RIS--assisted UWOC hop is modeled as:
\begin{equation}
    Z_u=\sum_{u=1}^{N}Y_u,
\end{equation}
where each $Y_u$ is an independent and identically distributed mEGG random variable incorporating underwater turbulence and pointing error impairments.

Following the first-term Laguerre series approximation  \cite{Primak2005}, the sum $Z_u$ can be accurately approximated by another random variable of the same family with parameters derived via the moment-matching method:
\begin{equation}
    \mathbb{E}\left(Z_u\right)=N\mathbb{E}\left(Y_u\right), \qquad \mathrm{Var}\left(Z_u\right)=N\mathrm{Var}\left(Y_u\right).
\end{equation}

Consequently, the shape and scale parameters in Eqs.~(21)--(22) of \cite{10413214} are given by:
{\small
\begin{equation}
    m_u=\frac{\left(\mathbb{E}\left(Z_u\right)\right)^2}{\mathrm{Var}\left(Z_u\right)}=N\frac{\left(\mathbb{E}\left(Y_u\right)\right)^2}{\mathrm{Var}\left(Y_u\right)},
    w_u=\frac{\mathrm{Var}\left(Z_u\right)}{\mathbb{E}\left(Z_u\right)}=\frac{\mathrm{Var}\left(Y_u\right)}{\mathbb{E}\left(Y_u\right)}.
\end{equation} }

This formulation indicates that the parameter $m_u$ exhibits a linear growth with respect to $N$, thereby mitigating the relative variance of the received signal and yielding a more tightly concentrated SNR distribution. Consequently, an increase in the number of O-RIS reflecting elements leads to a notable enhancement in system performance, as it directly improves the outage probability, average bit error rate, and average channel capacity of the UWOC link. These improvements are further substantiated by the analytical derivations presented in the subsequent sections.

\section{Appendix G \\ Evaluation of Passive and Active O-RIS Configurations}
Furthermore, unlike prior studies that often focus solely on the ideal passive O-RIS scenario, this Letter, for the first time, adopts a more comprehensive approach, investigating and evaluating various configurations of both passive and active O-RIS. This analysis is conducted using Monte Carlo simulations. By doing so, we not only assess system performance in an ideal scenario (which may be unattainable in the real world), but also consider the impact of various factors such as lack of control, phase error, quantization, and active noise on system performance. This approach provides us with a deeper understanding of O-RIS performance under different conditions. The following are the details and mathematical formulas for each of the five scenarios examined in this study.
\subsection{Passive Random O-RIS}
In this configuration, an O-RIS is present, but no control is exerted over its elements. The phase and amplitude of each element are determined completely randomly and independently of the channel. This scenario serves as a baseline, representing a situation where the O-RIS acts merely as a reflective surface with random properties, and no effort is made to optimize the signal. In the simulation, the phases and amplitudes are randomized once at the beginning of each Monte Carlo run for a specific SNR and remain fixed.
The SNR formula in this case is:
\begin{align}
\gamma_{_\text{Passive\_Random}} &= \gamma_{\text{avg}} \left| h_p \right|^2 \left| \sum_{m=1}^{N} \left( \rho_{_{m}} e^{j \theta_{m}} \right) h_{\alpha,m} \right|^2,
\end{align}
where
\begin{align*}
   & \rho_m \sim U(0,1) ,\quad \text{Uniform distribution between 0 and 1}, \\
& \theta_m \sim U(0,2\pi) , \quad \text{Uniform distribution between 0 and $2\pi$ }.
\end{align*}

\subsection{Passive Ideal O-RIS -- Fixed Zero Phase}

In this scenario, the O-RIS is modeled as a passive and lossless reflecting surface that perfectly compensates for the channel phases of the reflected paths. Each reflecting element applies an optimized phase shift such that the resulting effective phase of every reflected signal at the receiver is aligned to zero, ensuring coherent combination.
Let $h_{\alpha,m}$ denote the complex channel coefficient of the source--O-RIS--destination link via the $m$-th reflecting element. The O-RIS applies a phase shift $\theta_m$ that cancels the phase of this channel, i.e.,
\begin{equation*}
\theta_{m, Ideal} = -\angle(h_{\alpha,m}) = - (\angle h_{\alpha_{sr}} + \angle h_{\alpha_{rd}}) ) = -(\theta_{sr} +\theta_{rd}) .
\end{equation*}

As a result, the effective phase after RIS adjustment is
\begin{equation*}
\angle(h_{\alpha,m}) + \theta_m = 0,
\end{equation*}
which means that all reflected signals add up coherently at the receiver. This configuration guarantees maximum array gain and SNR enhancement. Since the reflection amplitude is assumed to be unity ($\rho_{_{m, Ideal}} = 1$), this model represents the upper performance bound of a passive O-RIS.
The instantaneous SNR for this case is given by

\begin{equation}
\begin{aligned}
\gamma&_{_{{Passive\_Ideal}}}
 = \gamma_{\text{avg}} |h_p|^2
\left| \sum_{m=1}^N \rho_{_{m, Ideal}} .h_{\alpha,m} . e^{j(\theta_{m,Ideal})} \right|^2 \\
&= \gamma_{\text{avg}} |h_p|^2\left| \sum_{m=1}^N |h_{\alpha,m}|. e^{j(\angle h_{\alpha,m})}.e^{j(\theta_{m, Ideal})}\right|^2 \\ &\hspace{2cm}=\gamma_{\text{avg}} |h_p|^2
\left| \sum_{m=1}^N |h_{\alpha,m}| \right|^2
\end{aligned}
\end{equation}

\subsection{Passive Controlled O-RIS -- With Phase Error}

This configuration represents a controllable passive O-RIS that attempts to adjust the signal phases to align them at the receiver, but this control is subject to random errors. The applied phase is set to compensate for the phase of the incoming channel (\(h_{\alpha,m}\)), i.e.,
$
\theta_m = -\angle \left(h_{\alpha,m}\right).
$
However, a random Gaussian error \(\epsilon_m\) is added, representing inaccuracies in Channel State Information (CSI) estimation or hardware implementation. The amplitude of each element is still assumed to be random, indicating a lack of precise amplitude control or noise. The resulting SNR for this configuration can be expressed as:
\begin{align}
\gamma_{_\text{Passive\_Controlled}} &= \gamma_{\text{avg}} \left| h_p \right|^2 \left| \sum_{m=1}^{N} \left( \rho_m e^{j(\theta_{m,Controlled})} \right) h_{\alpha,m} \right|^2.
\end{align}
where 
$
\rho_m \sim U(0,1),
\theta_{m,Controlled} = -\angle(h_{\alpha,m}) + \epsilon_m, 
\epsilon_m \sim \mathcal{N}(0, \sigma_{\text{phase\_error}}^2).
$

\subsection{Active Controlled O-RIS}

This is the most advanced O-RIS configuration. In addition to phase and amplitude control, it can amplify the received signals. Phase control in this case is assumed to be more precise with less error than in the passive-controlled case. However, active amplifiers add their own thermal noise to the signal, which must be accounted for in the final SNR calculation.
The SNR formula in this case is:

{ \small
\begin{align}
\gamma_{\text{Active\_Controlled}} &= \frac{\gamma_{\text{avg}} \left| h_p \right|^2 \left| \sum_{m=1}^{N} \left( G_{\text{active}} e^{j(\theta_{m,Active})} \right) h_{\alpha,m} \right|^2}{1 + N \cdot P_{\text{noise\_active\_element}}}
\end{align}}

\noindent ,where $\rho_m = G_{\text{active}}$, $\theta_{m,Active} = -\angle(h_{\alpha,m}) + \epsilon_m^\prime$, 
$\epsilon_m^\prime \sim \mathcal{N}(0, \sigma_{\text{phase\_error\_active}}^2)$, with a smaller variance, and 
$P_{\text{noise\_active\_element}}$ denotes the noise introduced by the active elements, which is added to the denominator of the SNR and has a non-zero value in the active case.

\subsection{Passive Quantized Phase O-RIS}

This scenario offers a more realistic model of a controllable passive O-RIS. Here, the O-RIS attempts to optimize the phase 
$
\theta_m = -\angle\left(h_{\alpha,m}\right),
$
but can only do so using a limited, discrete set of phase shifts (e.g., 1, 2, or 3 bits of quantization). This limitation leads to a quantization error, which reduces performance compared to the ideal passive case. The amplitude of each element is still assumed to be random.  The SNR formula in this case is:

{  \small
\begin{align}
\gamma_{\text{Passive\_Quantized}} &= \gamma_{\text{avg}} \left| h_p \right|^2 \left| \sum_{m=1}^{N} \left( \rho_{_m} e^{j \, \theta_{m,Quantized}} \right) h_{\alpha,m} \right|^2,
\end{align}}
where
\vspace{-2mm}
\begin{align*}
&\hspace{5mm} \rho_m \sim U(0,1), \\[6pt]
\theta&_{m,Quantized} = \text{Quantize}\!\left(-\angle(h_{\alpha,m})\right), \\[2pt]
Quant&ize(\phi) = 
    \text{round}\!\left(
        \frac{\operatorname{mod}(\phi, 2\pi)}{\tfrac{2\pi}{2^{\text{num\_bits}}}}
    \right) 
    \cdot \frac{2\pi}{2^{\text{num\_bits}}}.
\end{align*}

\vspace{5mm}

{\tiny
\bibliographystyle{IEEEtran}
\bibliography{REF1}
}

\vfill

\end{document}